# The Overstated Cost of AI Fairness in Criminal Justice

IGNACIO COFONE* & WARUT KHERN-AM-NUAI**

*The dominant critique of algorithmic fairness in AI decision-making, particularly in criminal justice, is that increasing fairness reduces the accuracy of predictions, thereby imposing a cost on society. This Article challenges that assumption by empirically analyzing the COMPAS algorithm, a widely used and widely discussed risk assessment tool in the U.S. criminal justice system.*

*This Article makes two contributions. First, it demonstrates that widely used AI models do more than replicate existing biases—they exacerbate them. Using causal inference methods, we show that racial bias is not only present in the COMPAS dataset but also worsened by AI models such as COMPAS. This finding has implications for legal scholarship and policymaking, as it (a) challenges the assumption that AI can offer an objective or neutral improvement over human decision-making and (b) provides counterevidence to the idea that AI merely mirrors preexisting human biases.*

*Second, this Article reframes the debate over the cost of fairness in algorithmic decision-making for criminal justice. It shows that applying fairness constraints does not necessarily lead to a cost in terms of loss in predictive accuracy regarding recidivism. AI systems operationalize concepts such as risk by making implicit and often flawed normative choices about what to predict and how to predict it. The claim that fair AI models decrease accuracy assumes that the model's prediction is an optimal baseline. Fairness constraints, in fact, can correct distortions introduced by biased outcome variables—which magnify systemic racial disparities in rearrest data rather than reflect actual risk. In some cases, interventions can introduce algorithmic fairness without imposing the cost often presumed in policy discussions.*

*These findings are consequential beyond criminal justice. Similar dynamics exist in AI-driven decision-making in lending, hiring, and housing, where biased outcome variables reinforce systemic inequalities beyond the choices of proxies. By providing empirical evidence that fairness constraints can improve rather than undermine decision-making, this Article advances the conversation on how law and policy should approach AI bias, particularly when algorithmic decisions affect fundamental rights.*

---

* Professor of Law and Regulation of AI, University of Oxford, Faculty of Law and Institute for Ethics in AI; Affiliated Fellow, Yale Law School Information Society Project. We thank Nikita Aggarwal, Jane Bailey, Mihailis Diamantis, Jim Gibson, Thomas Kadri, Orin Kerr, Kirsten Martin, Itay Ravid, Christopher Slobogin, Alicia Solow-Niederman, Rob Spear, Katherine Strandburg, Diane Uchimiya, Christopher Yoo, and Angela Zorro Medina for their feedback. The paper also benefited from comments received at the Privacy Law Scholars Conference and the Law & Society Conference. We thank Aya Amer, Wisaal Jahangir, and Bei Qi Zhou for their extraordinary research assistance, as well as the editors of the *ILJ*, in particular Paige Wynkoop, for their helpful comments and edits.

** Associate Professor and Desautels Scholar, Department of Information Systems, McGill University, Desautels Faculty of Management.

## INTRODUCTION

Discrimination in artificial intelligence (AI) decision-making is at the forefront of public discourse for good reason.[1] Decision-makers use AI support systems in criminal justice, chiefly to predict the probability that an individual will recommit a crime.[2] And these systems reveal stark racial biases. Methods to mitigate such biases have been criticized for increasing errors. Yet that criticism misses a crucial point: When the way to measure errors is biased, error rates are a bad metric for the cost of fairness.

---

1. *See, e.g.*, Rebecca C. Hetey & Jennifer L. Eberhardt, *The Numbers Don't Speak for Themselves: Racial Disparities and the Persistence of Inequality in the Criminal Justice System*, 27 CURRENT DIRECTIONS PSYCH. SCI. 183 (2018); Radley Balko, Opinion, *There's Overwhelming Evidence That the Criminal Justice System Is Racist. Here's the Proof.*, WASH. POST (June 10, 2020), https://www.washingtonpost.com/graphics/2020/opinions/systemic-racism-police-evidence-criminal-justice-system [https://perma.cc/A6T2-VYE6].

2. Sandra G. Mayson, *Dangerous Defendants*, 127 YALE L.J. 490, 493 (2018) ("It is hard to overstate the momentum behind this shift. . . . Jurisdictions around the country are increasingly turning to risk assessment as the keystone of pretrial reform."); *see also* Malathi A. & S. Santhosh Baboo, *An Enhanced Algorithm to Predict a Future Crime Using Data Mining*, 21 INT'L J. COMPUT. APPLICATIONS, May 2011, at 1, 5 ("Experimental results prove that the tool is effective in terms of analysis speed, identifying common crime patterns and future prediction.").

This Article makes two contributions. First, it provides empirical evidence that logistic regression algorithms, such as the *Correctional Offender Management Profiling for Alternative Sanctions* (COMPAS) algorithm, do not merely replicate racial biases that exist in their database, but *worsen* them.[3] While a discriminatory dynamic is often correctly attributed to bias in law enforcement,[4] we show that a source of bias amplification lies within the decision support systems themselves.

Second, it uses these empirical findings as a case study to uncover the flaw in the idea that AI fairness is socially costly. Risk assessment algorithms do not just predict outcomes; they also redefine accuracy in ways that decision-makers rarely question. It is possible for fairness tools to create a system that is not only fairer from a procedural and a substantive point of view, but also more accurate with regards to the prediction of violent criminality.

When AI systems are used, abundant empirical evidence shows that white applicants are more likely to be granted bail and parole than their Black counterparts, even when other factors captured by the AI system are similar, such as criminal charges and history.[5] This issue, while particularly prominent in criminal justice risk assessments, is broader. Scholars across data science and law show that biases in training data lead to discrimination and skew predictions of risk in several fields.[6] For example, data in credit-scoring processes may lead to discrimination and disparity-induced economic inequality.[7] Similarly, algorithms used to select tenants affect Hispanic people in a way that ultimately contributes to housing discrimination.[8] These examples show that it is important not only to detect AI biases but also to better understand the underlying mechanisms that drive them, so as to identify mitigation options.

---

3. To show this, we confirm racial bias in the COMPAS database by using causal tools that overcome critiques over prior findings.

4. *See* Emma Cunliffe, Opinion, *Let's Not Whitewash Racism in the Justice System*, EDMONTON J. (Oct. 24, 2018), https://edmontonjournal.com/opinion/columnists/opinion-lets-not-whitewash-racism-in-the-justice-system [https://perma.cc/5GUW-T4SX]; Ashley Hackett, *How Can U.S. Policymakers Fix the Broken Criminal Justice System?*, PAC. STANDARD (Apr. 25, 2018), https://psmag.com/social-justice/how-can-us-policymakers-fix-the-broken-criminal-justice-system [https://perma.cc/W3ZZ-MBCY].

5. *See, e.g.*, Jeff Larson, Surya Mattu, Lauren Kirchner & Julia Angwin, *How We Analyzed the COMPAS Recidivism Algorithm*, PROPUBLICA (May 23, 2016), https://www.propublica.org/article/how-we-analyzed-the-compas-recidivism-algorithm [https://perma.cc/YY6D-RX3F].

6. *See, e.g.*, Nikita Aggarwal, *The Norms of Algorithmic Credit Scoring*, 80 CAMBRIDGE L.J. 42 (2021); Jeremias Adams-Prassl, *What if Your Boss Was an Algorithm? Economic Incentives, Legal Challenges, and the Rise of Artificial Intelligence at Work*, 41 COMPAR. LAB. L. & POL'Y J. 123 (2019); Pauline T. Kim, *Data-Driven Discrimination at Work*, 58 WM. & MARY L. REV. 857 (2017) [hereinafter Kim, *Data-Driven Discrimination at Work*]; Anya E.R. Prince & Daniel Schwarcz, *Proxy Discrimination in the Age of Artificial Intelligence and Big Data*, 105 IOWA L. REV. 1257 (2020); Joseph Blass, Note, *Algorithmic Advertising Discrimination*, 114 NW. U. L. REV. 415 (2019); Sandra G. Mayson, *Bias In, Bias Out*, 128 YALE L.J. 2218, 2225 (2019).

7. Maddalena Favaretto, Eva De Clercq & Bernice Simone Elger, *Big Data and Discrimination: Perils, Promises and Solutions. A Systematic Review*, 6. J. BIG DATA, Feb. 2019, at 1, 5–7, 10; Aggarwal, *supra* note 6, at 55 ("[A] lender . . . could use its insights about consumers' biases and preferences to engage in less desirable forms of price discrimination and targeted marketing based on consumers' misperceptions . . . .").

Data scientists, as a response, have developed fairness-adjusted algorithms building on techniques to detect bias, including bias in the training dataset and bias in the algorithmic prediction process.[9] Algorithmic fairness, often defined as similar individuals or groups receiving similar treatments or outcomes,[10] has consequently drawn increasing attention from researchers and practitioners. Fairness-adjusted AI algorithms impose a sort of boundary on the model's prediction (i.e., the model's output, such as a "high-risk" or "low-risk" determination) so its predictions satisfy a previously determined notion of nondiscrimination, reducing bias in the output.[11]

But many argue that AI fairness comes at the hefty social cost of algorithmic accuracy.[12] In the criminal justice context, that alleged cost would be releasing individuals who will go on to commit more crimes, harming members of the public.[13] This argument leads to a question that neither data science nor law can answer on their own: Do other individuals pay a cost for AI fairness? Would society be worse off if fair AI were used in criminal justice, as this argument suggests?

The short answer to these questions, we find, is no: The idea that increasing fairness in COMPAS has an accuracy cost is flawed. Framing the issue of algorithmic fairness as one in which benefits for some are externalized as costs to others (or to society) is convenient for those who want to resist fairness limitations, but it does not capture the whole picture.[14]

---

8. Valerie Schneider, *Locked Out by Big Data: How Big Data, Algorithms and Machine Learning May Undermine Housing Justice*, 52 COLUM. HUM. RTS. L. REV. 251, 282 (2020) ("[A] housing provider seeking to exclude Latinos could set its algorithm to penalize individuals in particular industries in which Latinos are disproportionately represented . . . .").

9. Ninareh Mehrabi, Fred Morstatter, Nripsuta Saxena, Kristina Lerman & Aram Galstyan, *A Survey on Bias and Fairness in Machine Learning*, 54 ACM COMPUTING SURVS., July 13, 2021, at 1, 3–10.

10. Rachel K.E. Bellamy et al., *AI Fairness 360: An Extensible Toolkit for Detecting, Understanding, and Mitigating Unwanted Algorithmic Bias*, 63 IBM J. RSCH. & DEV., July–Sept. 2019, at 1, 2 ("Group fairness is the goal of groups defined by protected attributes receiving similar treatments or outcomes. Individual fairness is the goal of similar individuals receiving similar treatments or outcomes." (emphasis omitted)).

11. Sam Corbett-Davies, Johann D. Gaebler, Hamed Nilforoshan, Ravi Shroff & Sharad Goel, *The Measure and Mismeasure of Fairness*, 24. J. MACH. LEARNING RSCH., Aug. 2023, at 1, 4 [hereinafter Corbett-Davies et al., *The Measure and Mismeasure of Fairness*] ("[A] plethora of formal fairness criteria have been proposed in the computer science community.").

12. Sam Corbett-Davies, Emma Pierson, Avi Feller, Sharad Goel & Aziz Huq, *Algorithmic Decision Making and the Cost of Fairness*, 2017 KDD '17: PROC. 23RD ACM SIGKDD INT'L CONFER. ON KNOWLEDGE DISCOVERY & DATA MINING 797 [hereinafter Corbett-Davies et al., *Algorithmic Decision Making and the Cost of Fairness*]; *see also* Corbett-Davies et al., *The Measure and Mismeasure of Fairness, supra* note 11, at 8 ("A popular class of fairness definitions requires that error rates (e.g., false positive and false negative rates) are equal across protected groups.").

13. *See, e.g.*, Alex Chohlas-Wood, Madison Coots, Henry Zhu, Emma Brunskill & Sharad Goel, *Learning to Be Fair: A Consequentialist Approach to Equitable Decision Making*, MGMT. SCI., Dec. 2024, at 1 [hereinafter Chohlas-Wood et al., *Learning to Be Fair*]; Corbett-Davies et al., *Algorithmic Decision Making and the Cost of Fairness*, *supra* note 12, at 802–03; Alex Chohlas-Wood, Madison Coots, Sharad Goel & Julian Nyarko, *Designing Equitable Algorithms*, 3 NATURE COMPUTATIONAL SCI. 601 (2023) [hereinafter Chohlas-Wood et al., *Designing Equitable Algorithms*].

To address this question, we empirically examine the implications of using fair AI models in the criminal justice system. We use the COMPAS dataset as a case study to uncover broader issues in AI decision-making.[15] We detect different biases in the COMPAS training dataset and the COMPAS predictions. While these data have been analyzed before, we do so with causal methods of bias detection. The causal method overcomes the criticisms made of others' work on COMPAS-driven discrimination. It also allows us to report a novel finding: distinguishing the proportion of the COMPAS bias that is attributable to the dataset from the proportion attributable to the model itself.

Based on that finding, we identify a way to mitigate both biases (from the dataset and from the model) in COMPAS by employing common computational fairness criteria that have been used in prior literature on COMPAS, which allows us to engage with that literature on its own terms. As a result, we find that the common statement that society "pays" for fairness by losing accuracy is mistaken.[16] These results are consequential for law and policy.

The next Part of this Article presents legal and computer science discussions regarding fairness criteria and risk assessments in criminal justice. Part II presents our empirical study using causal methods, finding that the racial bias present in the COMPAS dataset is *amplified* by AI regression models such as COMPAS. Part III uses our results on COMPAS as a case study by imposing algorithmic fairness constraints in a second empirical evaluation. Part IV uncovers the broader implications of these empirical findings, explaining that algorithmic fairness constraints do not necessarily lead to a loss in accuracy as to the desired information when the algorithm uses imprecise predictor variables. In the case of COMPAS, fairness constraints do not reduce the type of accuracy that matters.

---

14. *See* Emily Black, John Logan Koepke, Pauline T. Kim, Solon Barocas & Mingwei Hsu, *Less Discriminatory Algorithms*, 113 GEO. L.J. 53, 56 (2024) ("Recent work in computer science has established that there are almost always multiple possible models with equivalent accuracy for a given prediction problem—a phenomenon termed *model multiplicity*." (emphasis added)).

15. *COMPAS Recidivism Risk Score Data and Analysis*, PROPUBLICA, https://www.propublica.org/datastore/dataset/compas-recidivism-risk-score-data-and-analysis [https://perma.cc/Z5JL-WZNP].

16. Note that prediction implies forecast, not certainty. Even though the standard language is to say "predict," algorithms can only make predictions about the likelihood of a future event and cannot predict any event: Algorithms can determine the likelihood at present of someone to recidivate at a future time, but not whether they will.

## I. AI BIAS IN CRIMINAL JUSTICE

### *A. Predictive AI in Criminal Justice*

AI decision support systems have become prevalent in the criminal justice system. Courts and parole boards across the United States use recidivism prediction instruments to assess the probability of reoffense among criminal defendants.[17]

The use of prediction software to evaluate recidivism is a pressing policy concern. These predictions inform decisions in the criminal justice system that carry life-altering consequences. Judges use prediction software to assist in deciding whether to grant bail—and sometimes in sentencing.[18] In parole decisions, board members use such predictions to assist in their deliberation on parole requests.[19] The decision to impose harsher sentences or to deny parole can have a drastic impact on affected individuals' lives.

ProPublica journalists who investigated COMPAS made its training dataset available to the public.[20] With this dataset, they and scholars have shown that COMPAS often mistakenly predicts Black individuals to have a high risk of recidivism.[21] Even after controlling for other characteristics such as age, gender, and prior crimes, Black individuals are given risk scores (i.e., their likelihood of committing a crime) that are higher than the rest of the population with similar characteristics.[22] ProPublica showed, in sum, that COMPAS assigns Black individuals a negative outcome (i.e., longer probation periods or prison time) more often than it does for white individuals.[23] The concern over these findings is rightly amplified by the fact that COMPAS reaches a performance ceiling of around 65%, which is approximately as accurate as predictions made by people with no criminal justice expertise.[24]

---

17. Julia Dressel & Hany Farid, *The Accuracy, Fairness, and Limits of Predicting Recidivism*, 4 SCI. ADVANCES, Jan. 17, 2018, at 1, 1–5.

18. *See* Gabriel Lima, Nina Grgić-Hlača & Meeyoung Cha, *Human Perceptions on Moral Responsibility of AI: A Case Study in AI-Assisted Bail Decision-Making*, CHI '21: POCS. 2021 CHI CONF. HUM. FACTORS COMPUTING SYS. 235.

19. *AI in the Criminal Justice System*, ELEC. PRIV. INFO. CTR., https://epic.org/issues/ai/ai-in-the-criminal-justice-system/ [https://perma.cc/2NR8-NF4P].

20. *COMPAS Recidivism Risk Score Data and Analysis*, *supra* note 15.

21. Aria Khademi, Sanghack Lee, David Foley & Vasant Honavar, *Fairness in Algorithmic Decision Making: An Excursion Through the Lens of Causality*, PROC. OF 2019 WORLD WIDE WEB CONF. 2907 [hereinafter *Fairness in Algorithmic Decision Making*]; Aria Khademi & Vasant Honavar, *Algorithmic Bias in Recidivism Prediction: A Causal Perspective (Student Abstract)*, 34 PROCS. AAAI CONFER. ON A.I. 13,839, 13,839–40 (2020) [hereinafter *Algorithmic Bias in Recidivism Prediction*].

22. Larson et al., *supra* note 5.

23. *Id.* ("[B]lack defendants who did not recidivate over a two-year period were nearly twice as likely to be misclassified as higher risk compared to their white counterparts . . . .").

24. Dressel & Farid, *supra* note 17, at 3. Since the AI-based decision support system is typically used by groups such as judges or parole board members—who are predominantly white—the bias from the AI system is likely to be amplified.

Other scholars have critiqued the results of the ProPublica study. The statistical critiques of researchers' bias-identification on COMPAS center on two concerns.[25]

The first contention is the base rate used for fairness assessments. Critics point out that, when comparing false positive rates between racial groups (i.e., cases where the algorithm mistakenly identifies a low-risk individual as someone who will reoffend), the algorithm's predictions need to be evaluated in the context of the actual prevalence of recidivism in each group, not through comparing the group averages with each other.[26] Failing to do so, critics argue, can lead to misleading conclusions about bias.[27] Notably, Sam Corbett-Davies and others argue that a reason for the COMPAS disparities is that "a greater fraction of black defendants have relatively high-risk scores, in part because black defendants are more likely to have prior arrests, which is a stronger indicator of reoffending."[28]

The second contention is that focusing solely on false positive rates ignores the equally socially significant issue of false negatives (i.e., cases where the algorithm fails to identify high-risk individuals who will reoffend).[29] While many critics of algorithmic fairness do acknowledge that arrest data are biased,[30] they argue that "satisfying common definitions of fairness means one must in theory sacrifice some degree of public safety."[31]

Our results and analysis provide a way to overcome both critiques. Regarding the first, there is, to our knowledge, no empirical or criminological evidence to substantiate the presumption that Black individuals commit crimes at a significantly higher rate than others.[32] Moreover, we show that, with COMPAS, race is not merely correlated with higher risk scores but actually causes higher risk scores.[33] Regarding the second, work in data science about less discriminatory alternatives suggests that reducing false positives does not necessarily imply increasing false negatives.[34] As the discussion section explains, we agree and our results align with this reality.[35] As argued by Hoffmann et al., "The legal system should not cling to something that is out of date when, at the same time, better alternatives are available for certain constellations."[36]

---

25. *See* Jennifer L. Skeem & Christopher T. Lowenkamp, *Risk, Race, and Recidivism: Predictive Bias and Disparate Impact*, 54 CRIMINOLOGY 680 (2016).

26. WILLIAM DIETERICH, CHRISTINA MENDOZA & TIM BRENNAN, NORTEPOINTE INC. RSCH. DEP'T, COMPAS RISK SCALES: DEMONSTRATING ACCURACY EQUITY AND PREDICTIVE PARITY 8 (2016), https://go.volarisgroup.com/rs/430-MBX-989/images/ProPublica_Commentary_Final_070616.pdf [https://perma.cc/NU8C-DTXC]; Megan Stevenson, *Assessing Risk Assessment in Action*, 103 MINN. L. REV. 303, 329–31 (2018) ("Disparate false positive rates are not a measure of racial bias under the definition used in this Article. Most other researchers do not measure racial bias using disparate false positive rates either."); *see also* Julia Angwin & Jeff Larson, *Bias in Criminal Risk Scores Is Mathematically Inevitable, Researchers Say*, PROPUBLICA (Dec. 30, 2016, 4:44 PM), https://www.propublica.org/article/bias-in-criminal-risk-scores-is-mathematically-inevitable-researchers-say [https://perma.cc/9XJZ-9J9Y]; Francesca Lagioia, Riccardo Rovatti & Giovanni Sartor, *Algorithmic Fairness Through Group Parities? The Case of COMPAS-SAPMOC*, 38 A.I. & SOC'Y 459 (2023).

27. Angwin & Larson, *supra* note 26; Lagioia et al., *supra* note 26, at 463–64.

28. Corbett-Davies et al., *Algorithmic Decision Making and the Cost of Fairness*, *supra* note 12, at 803.

29. *Id.*; DIETERICH ET AL., *supra* note 26, at 2–3.

30. Corbett-Davies et al., *Algorithmic Decision Making and the Cost of Fairness*, *supra* note 12, at 804–05.

31. *Id.* at 802; *see also* Mayson, *supra* note 6, at 2298–2300.

32. Ultimately, reliable quantitative data on who commits crimes does not exist—only data on who gets arrested does.

33. *See infra* Section I.C.

### *B. Fair(er) AI Algorithms*

From a sociotechnical perspective, one can classify machine learning biases that can lead to biased results into three types: bias in the process of building the model, bias in the sample that is used to train it, and social biases captured and perpetuated by the algorithm.

First, a biased process is a bias in how an algorithm processes information.[37] Biases in an algorithmic process often exist because human biases are translated into the system.[38] Even if no human chooses the outcome, there is human involvement in how that outcome is arrived upon: Humans frame the problem and make a choice about what the algorithm should predict.[39] Once that is decided, there is human involvement in selecting the features that will be used to generate the model's predictive outcome. Literature in psychology analyzes common sources of bias that are often translated into AI processes.[40]

Second, a biased sample exists when an algorithm is trained with a section of a dataset that is unrepresentative of the population, which will therefore produce nonrepresentative outputs.[41] Individual records, for example, may suffer from quality problems due to partial or incorrect data. The entire dataset might also have quality problems at higher rates for an entire protected group compared to others or it might be unrepresentative of the general population.[42]

---

34. Black et al., *supra* note 14, at 111.

35. *See infra* Section IV.A.

36. Hanna Hoffmann, Verena Vogt, Marc P. Hauer & Katharina Zweig, *Fairness by Awareness? On the Inclusion of Protected Features in Algorithmic Decisions*, 44 COMPUT. L. & SEC. REV. 1, Feb. 2022, at 1, 11.

37. Ignacio N. Cofone, *Algorithmic Discrimination Is an Information Problem*, 70 HASTINGS L.J. 1389, 1399–1402 (2019) ("The most evident type of algorithmic bias is a bias in the way in which an algorithm processes information: a bias in the model itself, or a classification bias. A biased process . . . .").

38. Batya Friedman & Helen Nissenbaum, *Bias in Computer Systems*, 14 ACM TRANSACTIONS ON INFO. SYS. 330, 333–36 (1996) (explaining the difference between preexisting bias and technical bias).

39. Pauline T. Kim, *Race-Aware Algorithms: Fairness, Nondiscrimination and Affirmative Action*, 110 CALIF. L. REV. 1539, 1548 (2022) [hereinafter Kim, *Race-Aware Algorithms*] ("[Bias] can also occur if the data encode human biases, such as supervisor evaluations of work performance or caseworker assessments of gang involvement that are shaped by implicit biases.").

40. Philip M. Podsakoff, Scott B. MacKenzie, Jeong-Yeon Lee & Nathan P. Podsakoff, *Common Method Biases in Behavioral Research: A Critical Review of the Literature and Recommended Remedies*, 88 J. APPLIED PSYCH. 879 (2003).

41. Kim, *Race-Aware Algorithms*, *supra* note 39, at 1548 ("Statistical bias can result when the data used to train the model are unrepresentative of the population or contain systematic errors."); Shira Mitchell, Eric Potash, Solon Barocas, Alexander D'Amour & Kristian Lum, *Algorithmic Fairness: Choices, Assumptions, and Definitions*, 8 ANN. REV.

Third, data can mirror social biases when an algorithm's training data are representative of the population but reflect and amplify prior systemic discrimination.[43] As Hanna Hoffmann explains, "A data set is always a reflection of the society from which the information has been obtained. If the society contains discriminatory elements and structures, these are also present in the training data set."[44] Consequently, despite being correctly trained with representative data, a machine learning process can still produce a disparate impact,[45] meaning a disproportionate adverse impact on a specific group,[46] because of embedded social inequalities.[47]

This Article's findings identify the source of bias in COMPAS's recommendations: a magnified bias of the third kind. That is, it identifies to what extent each of the three sources of bias is most dominant, leading to discrimination, as discussed below.[48] It then builds on work from computer science scholars who developed fairness-adjusted AI algorithms to alleviate bias, and eventually discrimination, in AI predictions, by applying one of these methods to COMPAS.[49]

---

STAT. & ITS APPLICATION 141 (2021) (discussing statistical bias).

42. Solon Barocas & Andrew D. Selbst, *Big Data's Disparate Impact*, 104 CALIF. L. REV. 671, 680–81, 684–87 (2016).

43. Kim, *Race-Aware Algorithms*, *supra* note 39, at 1548 ("[A]lgorithms may also produce skewed predictions because they reflect societal bias, accurately reproducing real differences between racial groups."); Cofone, *supra* note 37, at 1404–06; Mitchell et al., *supra* note 41 (discussing societal bias); *see* Harini Suresh & John Guttag, *A Framework for Understanding Sources of Harm Throughout the Machine Learning Life Cycle*, EEAMO 2021: ACM CONF. ON EQUITY & ACCESS ALGORITHMS, MECHANISMS, & OPTIMIZATION.

44. Hoffmann et al., *supra* note 36, at 12.

45. Alexandra Chouldechova, *Fair Prediction with Disparate Impact: A Study of Bias in Recidivism Prediction Instruments*, 5 BIG DATA 153, 154 (2017) ("Throughout our discussion we use the term *disparate impact* to refer to settings wherein a penalty policy has unintended disproportionate adverse impact on a particular group." (emphasis in original)).

46. Barocas & Selbst, *supra* note 42, at 673–74, 691.

47. Aylin Caliskan, Joanna J. Bryson & Arvind Narayanan, *Semantics Derived Automatically from Language Corpora Contain Human-Like Biases*, 356 SCI. 183 (2017); Daniel Rosenberg, *Data Before the Fact*, *in* "RAW DATA" IS AN OXYMORON 15 (Lisa Gitelman ed., 2013).

48. *See infra* Section I.C.

49. *See, e.g.*, Faisal Kamiran & Toon Calders, *Discrimination Aware Classification,* IEEE 2ND INT'L CONF. ON COMPUT., CONTROL & COMMC'N (2009); Indrė Žliobaitė, Faisal Kamiran & Toon Calders, *Handling Conditional Discrimination*, 11TH IEEE INT'L CONF. ON DATA MINING 992, 992–1001 (2011); Moritz Hardt, Eric Price & Nathan Srebro, *Equality of Opportunity in Supervised Learning*, NIPS' 16: PROCS. 30TH INT'L CONF. ON NEURAL INFO. PROCESSING SYS. 3323, 3324–30 (2016); Goce Ristanoski, Wei Liu & James Bailey, *Discrimination Aware Classification for Imbalanced Datasets*, PROCS. 22ND ACM INT'L CONF. ON INFO. & KNOWLEDGE MGMT. 1529, 1529–32 (2013); Benjamin Fish, Jeremy Kun & Ádám D. Lelkes, *Fair Boosting: A Case Study*, 2015 WORKSHOP ON FAIRNESS, ACCOUNTABILITY, & TRANSPARENCY MACH. LEARNING 1, 1–4; Muhammad Bilal Zafar, Isabel Valera, Manuel Gomez Rodriguez & Krishna P. Gummadi, *Fairness Constraints:*

These fair AI algorithms lead to common concerns for "losses" of information, which may arguably reduce accuracy in the outcome variable. Our results, as detailed below, show that the accuracy concerns should be tempered when examining the relationship between the target information (recidivism) and outcome variable (risk of arrest).[50] When one looks at the larger context, reduced accuracy for the outcome variable is not as concerning as it initially seems.[51] Notably, an aspect often missed in the conversation is that, because recidivism cannot be directly measured, rearrest is used to capture it.

### *C. Bias Identified Through Causal Methods*

Initial literature on COMPAS applied (noncausal) techniques that identify bias based on the disparities in prediction accuracy between subjects with different values of a sensitive variable (race). Notably, the ProPublica study used this method, running regressions on demographic factors (like race, age, criminal history, future recidivism, charge degree, and gender) to show that race is a predictive factor for a higher risk score.[52] This method of comparing error rates between groups to infer bias has been the subject of extensive critique.[53]

We find bias in COMPAS using causal methods that overcome such critiques.[54] The core of this causal analysis is going beyond identifying correlations to understand the effect of a variable (race) on the outcome. This task involves comparing outcomes with and without the cause while isolating other variables that might influence the result. We do so, as detailed below, for the COMPAS database and for model predictions.[55]

Causal methods are important for computational and normative reasons. Computationally, causal inference is crucial because correlations can lead to misleading conclusions, especially when it comes to statistical biases. For instance, a correlation might suggest that a certain group is less capable in a specific domain, but causal methods can reveal that the real issue is systemic bias, not inherent capability. Even if one found with a correlation method that Black individuals are assigned higher risk scores on average, it could be that Black individuals have

---

*Mechanisms for Fair Classification*, 54 PROCS. 20TH INT'L CONF. ON A.I. & STATS. 962, 962–69 (2017).

50. *See infra* Section IV.B.

51. *See infra* Section IV.C.

52. Larson et al., *supra* note 5.

53. Chouldechova, *supra* note 45, at 154–57; Alessandro Castelnovo, Riccardo Crupi, Greta Greco, Daniele Regoli, Ilaria Giuseppina Penco & Andrea Claudio Cosentini, *A Clarification of the Nuances in the Fairness Metrics Landscape*, 12 SCI. REPS., Mar. 10, 2022, at 1, 7–10; *see also* A. P. Dawid, *Causal Inference Without Counterfactuals*, 95 J. AM. STAT. ASS'N 407, 408–12 (2000) (explaining that counterfactual approaches require several assumptions about both the data and the mechanisms behind the effects evidenced in the data).

54. *See Algorithmic Bias in Recidivism Prediction*, *supra* note 21, at 13,839 (analyzing COMPAS with FACT); Maximilian Kasy & Rediet Abebe, *Fairness, Equality, and Power in Algorithmic Decision-Making*, FACCT '21: PROCS. 2021 ACM CONF. ON FAIRNESS, ACCOUNTABILITY & TRANSPARENCY 576, 582–83 (analyzing COMPAS with causal forest); *Fairness in Algorithmic Decision Making*, *supra* note 21 (establishing FACE and FACT).

55. *See infra* Part II.

different attributes (e.g., age) that influence the risk. We are, instead, interested in determining cause-and-effect relationships between variables. Normatively, in a context of bias, causal methods help identify systemic issues rather than individual instances. This advantage is essential for creating interventions that address the root causes of bias, rather than just its manifestations. Without understanding causal relationships, interventions might target symptoms rather than sources.

Our empirical study of the COMPAS dataset uses three widely recognized causal methods in the literature, detailed in the next Section, and adds the standard correlation study for comparison. The first two methods are called "Fair on Average Causal Effect" (FACE) and "Fair on Average Causal Effect on the Treated" (FACT).[56] The third method is an algorithm that estimates the average causal treatment effect called "causal forest."[57]

## II. Empirical Results on COMPAS's Double Bias

### *A. How We Causally Identified Bias in the COMPAS Dataset*

Our goal is to examine whether race *influences* the risk score. We use three methods to detect algorithmic bias: FACE, FACT, and causal forests. We use the COMPAS data collected by ProPublica, given the relevance of this dataset for prior data science studies and for its legal and policy ramifications.[58] This dataset is often used in research to demonstrate racial bias in AI predictions in the criminal justice system.[59]

Both FACE and FACT work by building logistic regressions and weighting sensitive variables, such as gender and race, to compare the output values (in the case of COMPAS, high risk and low risk) with the sensitive variables (in the case of COMPAS, race).[60] Both methods have been demonstrated to be reliable on multiple public datasets.[61] The term "Fair on Average," which they share, implies that the causal effect of a variable (race) is evaluated across groups to see that the average effect of the cause does not disproportionately benefit or harm any group. For example, in a healthcare setting, a treatment's effectiveness might be evaluated to ensure it is equally effective across different demographic groups. While FACE does this by focusing on fairness constraints in the population in general, FACT does so by focusing on fairness constraints in the protected group.[62] For example, FACE can be used to examine gender bias in pay by comparing the outcome variable (salary) between female and male employees as groups. Meanwhile, FACT can find the best-matching observations which differ only in the sensitive variables (gender); it would discover the matching pairs of female and male employees who have similar observable characteristics that may affect the outcome variable (e.g., educational background and experience) and compare their salary levels to ascertain whether gender bias exists.[63]

---

56. *Fairness in Algorithmic Decision Making*, *supra* note 21, at 2907.

57. Stefan Wager & Susan Athey, *Estimation and Inference of Heterogeneous Treatment Effects Using Random Forests*, 113 J. AM. STAT. ASS'N 1228, 1229 (2018) [hereinafter *Estimation and Inference of Heterogeneous Treatment Effects*] ("[O]ur proposed forest is composed of *causal trees* that estimate the effect of the treatment at the leaves of the trees; we thus refer to our algorithm as a *causal forest*." (emphasis in original)); Susan Athey & Stefan Wager, *Estimating Treatment Effects with Causal Forests: An Application*, 5 OBSERVATIONAL STUD. 37 (2019) [hereinafter *Estimating Treatment Effects with Causal Forests*].

58. *See infra* Appendix A.

59. Larson et al., *supra* note 5; *Algorithmic Bias in Recidivism Prediction*, *supra* note 21, at 13,839–40.

60. Hajime Shimao, Warut Khern-am-nuai, Karthik Kannan & Maxime C. Cohen, *Strategic Best Response Fairness in Fair Machine Learning*, AIES '22: PROCS. 2022 AAAI/ACM CONF. ON A.I., ETHIC & SOC'Y 664, 672, 675.

61. *Algorithmic Bias in Recidivism Prediction*, *supra* note 21, at 13,839–40 (these include the adult income dataset from the UCI repository, NYC Stop and Frisk data, and the COMPAS dataset with recidivism prediction).

FACE is a weighting method—it uses inverse probability weighting.[64] For each observation in our dataset, we assign a weight from one to infinity to make the control group and the treatment group similar. We add a larger weight to individuals who are underrepresented in the sample and a lower weight to those who are overrepresented (e.g., if the population is 50% female, but the sample is only 35% female, a larger weight would be added to female observations). FACE accounts for confounding variables. For example, suppose Black individuals are on average younger than other individuals in the dataset. Such a discrepancy could explain why Black individuals have higher risk scores than other individuals. FACE accounts for this discrepancy by down-weighting young Black individuals in the estimation process and up-weighting young non-Black individuals. With this approach, characteristics other than the sensitive variable are equivalent between the group of Black and of non-Black individuals.

FACT is a matching method. The objective of FACT is to make the control and treatment groups similar.[65] So, instead of weighting, FACT discards observations that are dissimilar across the control group and the treatment group. For example, if there are more young Black individuals than young white individuals in the dataset, FACT will remove young Black individuals from the dataset (assigning them a weight of zero) until group ages are balanced. We use three matching methods to assess the similarities between Black individuals and other individuals in the COMPAS dataset under FACT.[66]

The third method is a causal forest: a machine learning algorithm used for understanding the underlying mechanisms that drive a machine learning output based on causal inference.[67] Causal forests leverage a machine learning model called random forest,[68] which is widely used for classification and regression tasks.[69] The random forest model works by training a large number of decision trees on random subsets of the data and then combining the outputs of those trees to make a final prediction.[70] The advantage of random forests is that they can handle large datasets and are able to automatically, rather than manually, identify the most important features in the data. Additionally, they can generalize to new data (in technical terms, they are resistant to overfitting).[71]

---

62. Donald B. Rubin, *Causal Inference Using Potential Outcomes: Design, Modeling, Decisions*, 100 J. AM. STAT. ASS'N 322 (2005).

63. Therefore, under FACE the underlying groups may be different but one would still compare them with each other through weighting, while with FACT one selects similar individuals where the only difference is the protected category (i.e., race) to examine pairs of cases that are equivalent other than in their race.

64. Daniel E. Ho, Kosuke Imai, Gary King & Elizabeth A. Stuart, *Matching as Nonparametric Preprocessing for Reducing Model Dependence in Parametric Causal Inference*, 15 POL. ANALYSIS 199 (2007).

65. *Id.*

66. *See infra* Appendix B.

67. *Estimation and Inference of Heterogeneous Treatment Effects*, *supra* note 57, at 1232–48.

The difference between causal forests and random forests is that a causal forest is designed to identify and model causal relationships in the data, rather than just predicting the outcome of a given input like a random forest does. The causal forest model repeatedly splits the data and minimizes differences across splits regarding the relationship between the independent variable and the dependent variable of interest.[72] This process allows the analysis to approximate conditions similar to those of a quasi-randomized experiment.[73]

Since our objective is to test the effect of the sensitive variable (race) on the outcome variable (risk score for rearrest) by comparing individuals with similar characteristics, we first divide the dataset into a "treatment group" (Black individuals) and a "control group" (non-Black individuals, which includes all other groups, used as a baseline for accuracy). Then, we match individuals with similar characteristics into a series of categories.[74] We then compute the difference in the outcome variable between matched individuals across each category.[75] This way, all other predictors are controlled so we can causally estimate the impact of race on the outcome variable.

---

68. *Id.* at 1228 ("[W]e develop a nonparametric *causal forest* for estimating heterogeneous treatment effects that extends Breiman's widely used random forest algorithm." (emphasis in original)); *Estimating Treatment Effects with Causal Forests*, *supra* note 57, at 37, 39–41 ("Causal forests are an adaptation of the random forest algorithm of Breiman (2001) to the problem of heterogeneous treatment effect estimation.").

69. *See, e.g.*, Susan Athey & Guido Imbens, *Recursive Partitioning for Heterogeneous Causal Effects*, 113 PROC. NAT'L ACAD. SCI. 7353 (2016); Susan Athey, Julie Tibshirani & Stefan Wager, *Generalized Random Forests*, 47 ANNALS STATS. 1148 (2019); Julie Tibshirani, Susan Athey, Rina Friedberg, Vitor Hadad, David Hirshberg, Luke Miner, Erik Sverdrup, Stefan Wager & Marvin Wright, *Generalized Random Forests*, GRF 2.4.0, https://grf-labs.github.io/grf/index.html [https://perma.cc/VGD3-LHVY].

70. A decision tree is a traditional algorithm with flowchart-like structure where each node represents a test (such as a yes/no question), each branch represents the outcome of the test (such as if yes, then X), and each leaf node represents a label.

71. Hasan Ahmed Salman, Ali Kalakech & Amani Steiti, *Random Forest Algorithm Overview*, 2024 BABYLONIAN J. OF MACH. LEARNING 69, 78.

72. This process, which measures the average effect of a treatment variable for individuals with similar characteristics, is called Conditional Average Treatment Effects (CATE).

73. Athey & Imbens, *supra* note 69, at 7353 ("Within the prediction-based machine learning literature, regression trees differ from most other methods in that they produce a partition of the population according to covariates, whereby all units in a partition receive the same prediction.").

74. We build multiple branches to divide them into different categories. The trees are built in a way to maximize the differences across each category and minimize the differences within a category.

*B. Biased Results in the COMPAS Training Dataset*

We find racial bias in the COMPAS dataset with all three methods: FACE, FACT, and causal forest.[76] The estimated bias values range from 12% to 19%, depending on the method.[77] That means that the average Black individual receives a risk score (likelihood of reoffending) 12% to 19% higher than the rest of the population with *all things equal* (i.e., controlling for other characteristics, such as criminal history, socioeconomic status, and location). We also find bias estimates for Hispanic individuals between 5% and 9% higher than the rest of the population, also controlling for all other variables.[78] This evidence is stronger than analyses that show rates of positive and negative predictions because it is causal. Because our three methods (FACE, FACT, and causal forest) use different methodologies so that causal inferences are established in the estimation process, our results provide evidence that significant bias caused by race exists in the COMPAS dataset against Black and Hispanic individuals.[79]

For robustness, we also analyze the COMPAS dataset with the same traditional, correlation-based methods used by other researchers in the past, to compare their results with ours. The correlational method detects the difference in prediction accuracy between predictions of Black individuals and of non-Black individuals—as opposed to establishing the impact of race on risk score as the causal methods do. We find racial bias in the dataset with this method as well.[80] Our results thus confirm findings from prior studies.[81]

Additionally, we find that a simple regression on the dataset performs better for non-racialized individuals than it does for Black and Hispanic individuals. In other words, Black and Hispanic individuals receive lower prediction accuracy. The accuracy of a logistic regression model is 89% when used for non-Black individuals, but only 81% when used for Black individuals.[82] We find a similar effect for non-Hispanic versus Hispanic individuals, with a drop of 88% to 82%.[83] This difference means that the COMPAS dataset is less informative for Black and Hispanic individuals than it is for others.[84]

---

75. We isolate the relationship between treatment and outcome variable with a residual-on-residual regression.

76. See *infra* Table 3 & Appendix B for details on our results.

77. *See infra* Appendix B, Table 3.

78. Although with a lower (still statistically significant) p-value, as illustrated in Appendix B. Table 3b, in Appendix B, presents equivalent results from the causal forest method.

79. Table 2, in Appendix A, provides summary statistics of variables in the COMPAS Dataset.

80. For this analysis, we use a logistics regression model with tenfold cross validation on the COMPAS dataset, similar to the one performed by ProPublica.

81. *Fairness in Algorithmic Decision Making*, *supra* note 21; Larson et al., *supra* note 5.

82. *See infra* Appendix B, Table 5.

83. *See infra* Appendix B, Table 5b.

The story, however, gets worse. In addition to the COMPAS dataset's negative impact on racialized individuals and its poor predictive performance for them, AI algorithms like COMPAS worsen the bias, as the next Section shows.

### *C. Common AI Models Increase Bias from the Training Dataset*

We investigate whether the bias that exists in the COMPAS training dataset changes when we use a logistic regression AI model. While the exact COMPAS model is protected by trade secrecy, it is well known that COMPAS closely resembles this type of regression model.[85]

We measure bias with a machine learning metric called "disparate impact ratio," which, relying on demographic parity, assesses whether a model's predictions disproportionately affect one group of individuals over another. It does so by comparing the percentage of favorable outcomes for a monitored group to the percentage of favorable outcomes for a reference group.[86]

We calculate the model's disparate impact ratio by dividing the ratio of a positive outcome for the protected group (Black individuals) by the ratio of a positive outcome for a majority group (all other individuals). Here, "positive" means "high risk," and "negative" means "low risk." Hence, a false positive is mistakenly classified as high risk and a false negative is mistakenly classified as low risk. As COMPAS applies a risk score of one to ten where one is the lowest risk in the scale and ten is the highest,[87] we define the outcome as low risk (negative prediction) for observations with a risk score from one to five and as high risk (positive prediction) for observations with a risk score from six to ten.

We first calculate the disparate impact ratio based on the original COMPAS dataset. Following that, we use the dataset as training data for a logistic regression model and calculate the disparate impact ratio of its predictions. A lower percentage in disparate impact ratio indicates a higher bias, as treating both groups the same would lead to a ratio of 1. We find that the disparate impact ratio in the original dataset is 0.48 while the disparate impact ratio in AI predictions is 0.37. We also find equivalent effects for Hispanic individuals.[88]

---

84. *See infra* Appendix B, Tables 5 (Black individuals), 5b (Hispanic individuals).

85. Logistic regression is one of the most commonly used classification models, both in AI and in statistics. The basic idea is to rely on a logit function to establish the relationship between predictors (x) and the categorical target variable (y). *See generally* DANIEL T. LAROSE & CHANTAL D. LAROSE, DATA MINING AND PREDICTIVE ANALYTICS §§ 3.1–3.12 (2d ed. 2015).

86. Bellamy et al., *supra* note 10, at 2.

87. Larson et al., *supra* note 5.

88. *See infra* Appendix B (noting lower p-values, still statistically significant, for most tests).

**Table 1:** Bias from AI Model Versus Bias in Training Dataset

| Category | Disparate Impact Ratio |
| --- | --- |
| Original Dataset | 0.4821 |
| AI Predictions | 0.3755 |

For context, the computer science literature often uses an "80% rule" as a cutoff when measuring fairness: If members of the protected group receive beneficial outcomes (true and false negatives) 80% of the time less than members of the majority group, that indicates an adverse impact under demographic parity.[89] Both of the disparity values we found, reported in Table 1, at 48% and 37%, are significantly below the 80% metric.

These results show that, when the COMPAS training dataset is used to train common AI models, the resulting models are more biased than the dataset was. Our analysis shows that the use of a basic AI model with the biased training dataset amplifies the dataset's racial disparity by 20.58%.[90] As a consequence, the algorithm increases Black individuals' likelihood to be classified as high risk compared to white individuals, called a risk ratio, by 28.5%.[91] While the test shows what common AI models do with database bias—rather than COMPAS specifically—one thing is worth noting: Since computer science literature has shown that these models perform *better* than COMPAS,[92] the COMPAS model likely increases bias over its database by *at least* this amount.

These results expand (and add nuance to) foundational work that shows that AI *replicates* social bias.[93] Our results illustrate that common AI algorithms do more than replicate social biases in their output and perpetuate them when broadly deployed: They also *amplify* those biases.

---

89. Castelnovo, *supra* note 53, at 1, 6; Xi Xin & Fei Huang, *Antidiscrimination Insurance Pricing: Regulations, Fairness Criteria, and Models*, 28 N. AM. ACTUARIAL J. 285, 296 (2023) (indicating the widespread use of the 80% cutoff and explaining that it was originally taken from the 1978 Uniform Guidelines for Employee Selection Procedures). In other words, the 80% Rule states that the selection rate for a protected group should be at least four-fifths of the selection rate for the majority or reference group.

90. The 10.66 percentage points difference represents a 20.58% increase: $((1-0.3755)-(1-0.4821))/(1-0.4821)\times100$. Such a difference is statistically significant at a p-value $< 0.001$ based on the permutation test with 10,000 iterations. Detailed results are reported in Table 6 in Appendix B.

91. In practice, the 20.58% increase in disparate impact ratios means that Black individuals are 2.07 times more likely to be classified as high risk than white individuals relative to their population sizes under the dataset disparity value of 0.4821, while they are 2.66 times more likely to be classified as high risk under the algorithm's disparity value of 0.3755.

92. Chouldechova, *supra* note 45.

93. *See, e.g.*, SHANNON VALLOR, THE AI MIRROR: HOW TO RECLAIM OUR HUMANITY IN AN AGE OF MACHINE THINKING 15–35 (2024); SAFIYA UMOJA NOBLE, ALGORITHMS OF OPPRESSION: HOW SEARCH ENGINES REINFORCE RACISM 171–81 (2018); *see also* Caliskan et al., *supra* note 47, at 183.

An intuitive explanation of why the AI model is more biased than its training dataset, detailed in the discussion below,[94] is as follows. The model is tasked to maximize accuracy for the outcome variable (risk). And, based on a biased predictor variable in the dataset (arrest), the model finds traits that correlate with being Black to be predictors of being at high risk of rearrest. This finding matters for generalizability: All AI models trained to maximize the accuracy of a target, when using a biased outcome variable (as rearrest is), will have this problem. Rather than being a problem of COMPAS, this is a problem of all AI models that aim to maximize prediction accuracy while working with units of measurements that incorporate social biases—which are most AI models currently used in decision-making. These models are trained to maximize the accuracy of a biased outcome variable whenever the target information is not subject to direct measurement.[95] We should thus expect most decision-making AI models to be worse in terms of bias than their training data when involving the third type of bias we define above.[96]

## III. The Case Study Is Informative of AI Fairness Generally

### *A. Mitigating AI Bias*

After we detect bias in the dataset and prediction results, we use a second methodology to mitigate the bias found in AI predictions.[97] This second approach manipulates the prediction process so that results (outputs) comply with a notion of fairness (nondiscrimination). We examine what happens when fairness-adjusted AI models designed to mitigate bias in AI prediction results are used on this dataset.

Choosing a fairness criterion is not uncontroversial. As Anne Washington explains, "There is no single mathematical definition of fairness. The people developing a 'fair' algorithm must decide on the uniformity or variation that is necessary for a functioning system. Data science experts conclude that the people who control the algorithms define fairness."[98] The lack of a uniform definition is a routine obstacle in the literature on AI fairness, transparency, and accountability reflected in law.[99] The fairness theories used are where many of the judgment calls are made.[100]

---

94. *See infra* Part IV.

95. For example, models trained to identify similarity with people who paid their loans in the past for mortgage allocation (when being a reliable borrower is unmeasurable) or similarity with people who currently live in a neighborhood for housing advertising (when being someone who would like to live there is unmeasurable) would also face this problem.

96. *See supra* Section I.B.

97. See *infra* Appendix B for details.

98. Anne L. Washington, *How to Argue with an Algorithm: Lessons from the COMPAS-ProPublica Debate*, 17 COLO. TECH. L.J. 131, 151 (2018) (footnote omitted).

99. *See* Kim, *Race-Aware Algorithms*, *supra* note 39, at 1549.

100. *See* Alex Hanna, Emily Denton, Razvan Amironesei, Andrew Smart & Hilary Nicole, *Lines of Sight*, LOGIC(S), Dec. 20, 2020, at 51, 63 ("As Gebru explained during a tutorial at the Computer Vision and Pattern Recognition conference in June 2020, 'Fairness is not just

We apply two measures of fairness in machine learning algorithms commonly used in situations where it is important to avoid discrimination based on sensitive attributes, such as race or gender: (1) demographic parity (also known as statistical parity) and (2) equalized odds. We chose these because they are two of the dominant fairness criteria used to evaluate machine learning algorithms in the data science literature.[101] We expect equivalent results for other definitions of fairness in the literature if the analysis is run for them.

Demographic parity requires that the beneficial predictions made by an algorithm not be systematically different for different demographic groups. In other words, it compares beneficial predictions against all predictions and proposes that the fraction of beneficial predictions should be equal across groups.[102] For example, if an algorithm is used to determine whether someone will be approved for a loan, demographic parity would require that the algorithm should have a similar rate of approved applicants for both demographic groups.

Equalized odds requires equivalent false positive and false negative rates for different demographic groups. For example, if an algorithm is used to predict whether someone will default on a loan, equalized odds would require that the algorithm does not have a higher false negative rate (i.e., failing to identify potential defaulters) for one demographic group compared to another. To satisfy the constraint, the error rates in classification should be the same for both groups.

The metrics are related. Demographic parity, on the one hand, requires that the selection rates with a beneficial outcome are equal across groups. In this scenario, an equivalent fraction of white individuals and Black individuals should be considered low risk. It requires that the *low-risk prediction* is *statistically* independent of race (i.e., the presence of the sensitive variable does not affect the probability of being classified as low risk).[103] Equalized odds, on the other hand, requires both true positive rate parity and false positive rate parity between groups. In this context, a classifier will satisfy equalized odds if the prediction (*high risk or low risk*) is *conditionally* independent of race (i.e., the sensitive variable and the prediction are independent of each other given other relevant conditions).[104]

---

about datasets, and it's not just about math. Fairness is about society as well, and as engineers, as scientists, we can't really shy away from that fact.'").

101. *See, e.g.*, Hardt et al., *supra* note 49, at 3323; Alekh Agarwal, Alina Beygelzimer, Miroslav Dudík & John Langford, *A Reductions Approach to Fair Classification*, 2017 FAT ML '17: PROCS. FAIRNESS, ACCOUNTABILITY & TRANSPARENCY MACH. LEARNING 1.

102. Kim, *Race-Aware Algorithms*, *supra* note 39, at 1577–78 ("At the other end of the spectrum are strategies aimed at ensuring proportional outcomes—what computer scientists refer to as 'demographic parity.' These strategies seek to equalize the probability of a positive outcome across demographic groups. Put differently, they ensure that demographic groups receive positive outcomes in proportion to their actual representation.").

103. Agarwal et al., *supra* note 101, at 2 ("A classifier h satisfies demographic parity under a distribution over (X, A, Y ) if its prediction h(X) is statistically independent of the protected attribute A . . . ." (emphasis omitted)).

104. Hoffmann et al., *supra* note 36, at 2 ("A system that does not judge the minority group significantly better or worse than the group that represents the majority is called fair. One possible fairness measure to meet this requirement is called Conditional Independence. According to this fairness measure, the system would be fair if the ratio of people invited to an interview out of the applicants considered suitable is the same for both groups." (footnote

### B. Examining Accuracy

We identified a model that satisfies the fairness constraint of demographic parity by adjusting model parameters (what data scientists call hyperparameter tuning).[105] Imposing such a fairness constraint ensures that the model predicts Black individuals to have a high risk of recidivism at a similar rate that it predicts it for white individuals.

To do so, we use the GridSearch method,[106] which operationalizes a popular black-box approach that modifies prediction outcomes by sequentially relabeling and reweighting.[107] Its idea is to improve the performance of a model by systematically searching through a range of values. Figure 1 plots the different fairness-adjusted models produced with this method based on two metrics: the model's fairness, as measured by the difference between the ratio of Black individuals with beneficial outcomes versus the ratio of non-Black individuals with beneficial outcomes, and the model's accuracy with respect to the outcome variable. We find a marked correlation: The smaller the selection rate difference, the lower the model's accuracy for the outcome variable.

Figure 1 plots these models based on selection rate difference and classification accuracy.

---

omitted)). *See also id.* ("A classifier h satisfies equalized odds under a distribution over (X, A, Y) if its prediction h(X) is conditionally independent of the protected attribute A given the label Y . . . ." (emphasis omitted)).

105. *See id.* at 6–7; James Bergstra, D. Yamins & D.D. Cox, *Making a Science of Model Search: Hyperparameter Optimization in Hundreds of Dimensions for Vision Architectures*, PROCS. 30TH INT'L CONF. ON MACH. LEARNING 115, 115 (2013) ("[T]uning of hyperparameters is an important part of understanding algorithm performance, and should be a formal and quantified part of model evaluation.").

106. *See* SARAH BIRD, MIROSLAV DUDÍK, HANNA WALLACH & KATHLEEN WALKER, FAIRLEARN: A TOOLKIT FOR ASSESSING AND IMPROVING FAIRNESS IN AI (2020); s*ee also* fairlearn.reductions.GridSearch, FAIRLEARN, https://fairlearn.org/main/api_reference/generated/fairlearn.reductions.GridSearch.html [https://perma.cc/5TVW-NXVJ] (providing details of the package); *GridSearch with Census Data*, FAIRLEARN, https://fairlearn.org/main/auto_examples/plot_grid_search_census.html [https://perma.cc/RY3E-ZP87] (providing an example of how the package can be used).

107. Agarwal et al., *supra* note 101, at 60 ("The second group of approaches eliminate the restriction to specific classifier families and treat the underlying classification method as a 'black box,' while implementing a wrapper that either works by pre-processing the data or post-processing the classifier's predictions . . . .").

**Figure 1:** Model Accuracy and Selection Rate Difference (Grid Search with Demographic Parity)

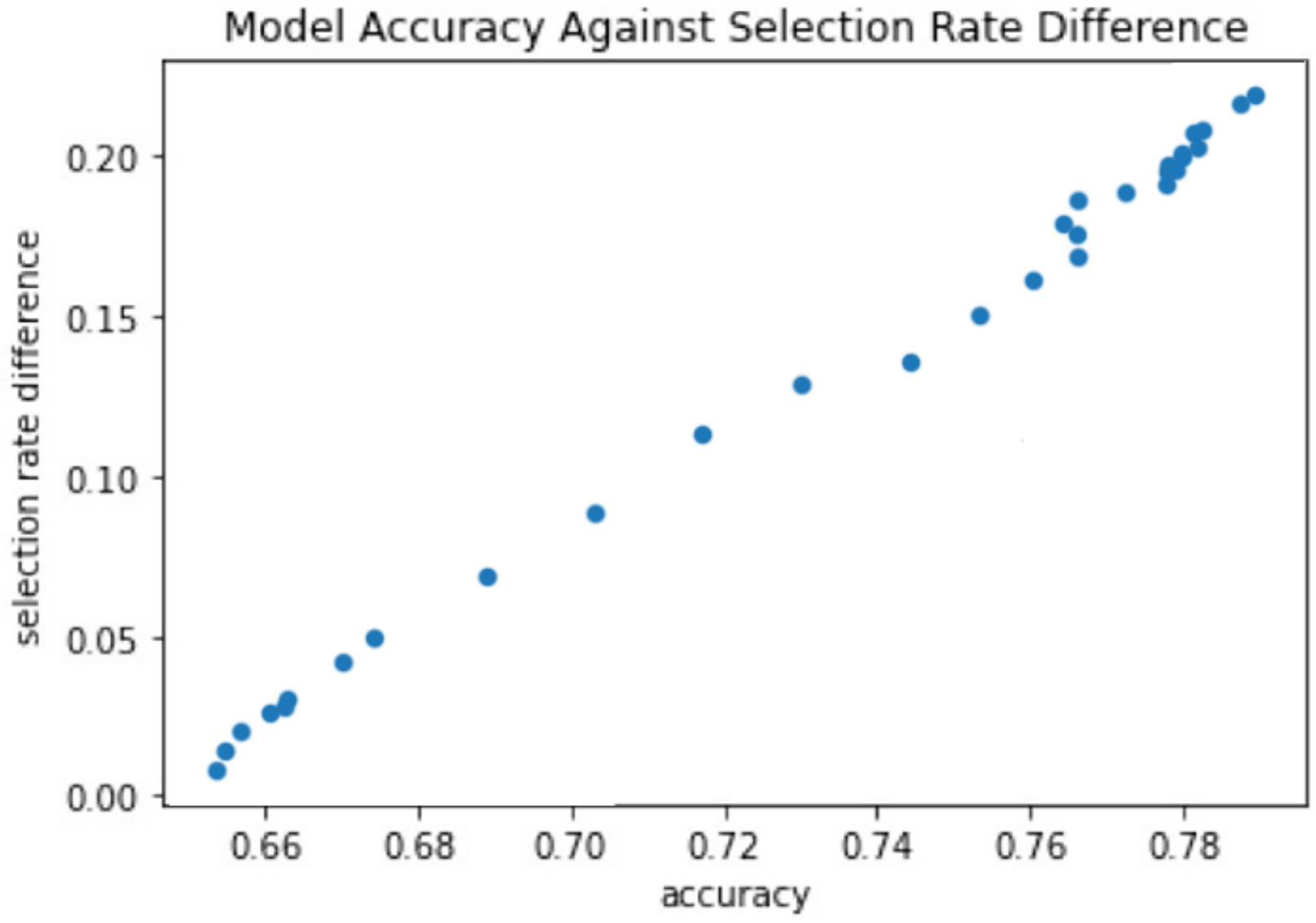

Model 0 (top right corner) represents the usual model that maximizes classification accuracy. Model 29 (bottom left corner) has a selection rate difference close to zero. The selection rate difference, on the *y* axis, measures the absolute difference in the rate at which a beneficial decision outcome is assigned for each of the two groups, while the disparate impact ratio compares them by calculating the ratio of the selection rates.[108]

We take the standard model that maximizes accuracy (shown in the top right corner) as a benchmark for comparison against fairness-adjusted models. Comparing this standard model with one that was most adjusted to achieve a fair outcome, with a selection rate difference across groups close to zero (shown in the bottom left corner), one can see that the "accuracy" of the fairness-adjusted model with regards to the outcome variable is low. We find losses in accuracy of 30.93% for Black individuals and of 4.51% for non-Black individuals. And we find losses in accuracy of 22.08% for Hispanic individuals and of 4.09% for non-Hispanic individuals. The classification accuracy of these models based on Black individuals and non-Black individuals, which is detailed in the Appendix, seems to point to a correlation between selection rate difference and accuracy with regards to the outcome variable.[109] These data seem to point, in sum, to a social "cost" of fairness (measured as selection rate difference) in terms of model accuracy.

The fair AI models seem to especially sacrifice model accuracy when incorporating fairness for members of the protected groups (Black and Hispanic

108. *See supra* Section II.C. (discussing the disparate impact ratio).
109. *See infra* Appendix B, Tables 7, 7b.

individuals). In other words, the fair models seem more likely to "overlook" a prediction that a Black or Hispanic individual has a high risk of recidivism (rearrest) so as to force the selection rate to be equal across two groups. As a result, it might appear that fairness decreases model accuracy for Black and Hispanic individuals significantly.

The original dataset alone cannot answer this question because it cannot tell us which individuals reoffended. Hence, we use a supplemental dataset from a follow-up study that contains information on whether each individual in the original dataset reoffends in the two years after being released from jail.[110] Note that we process the dataset to alleviate a preprocessing issue that occurs in the original dataset.[111] Using this dataset, based on prediction results from model 0 and model 29, we calculate a confusion matrix.[112]

The number of false-negative predictions decreases drastically for all groups. For Black individuals, it decreases from 63 to 2 (96.83% decrease), and for non-Black individuals from 27 to 1 (96.30% decrease).[113] It similarly decreases from 67 to 7 (89.55% decrease) for Hispanic individuals and from 23 to 4 for non-Hispanic individuals (82.61% decrease).[114] As a result, individuals who have low recidivism risk are extremely unlikely to be predicted to have high risk.

False-positive predictions, however, increase. They increase from 107 to 250 (133.64% increase) for Black individuals, from 124 to 171 for non-Black individuals (37.90% increase), from 126 to 251 for Hispanic individuals (99.21% increase), and from 105 to 145 for non-Hispanic individuals (38.10% increase).[115] Note that the percentage reductions in false negatives are close to 100% for all groups, and the increments in false positive rates are comparable for racialized groups while lower for non-racialized groups.

Because the results from this analysis are bound by the fairness constraint we chose (demographic parity), we repeated the analyses using a similar algorithm with equalized odds as the fairness constraint. We attained equivalent results.[116]

### *C. Mortgage Loans: Equivalent Results with Different Data*

The implications of our results are broader than risk assessment algorithms and than AI fairness in criminal law. Credit, for example, has an equivalent problem.[117]

---

110. *COMPAS Recidivism Risk Score Data and Analysis*, *supra* note 15.

111. Matias Barenstein, ProPublica's COMPAS Data Revisited 1, 1–2 (June 13, 2019) (unpublished manuscript) (on file with arXiv), https://arxiv.org/pdf/1906.04711 [https://perma.cc/89U5-HGNS] ("ProPublica made an important data processing error when it created these datasets, failing to implement a two-year sample cutoff rule for recidivists in such datasets (whereas it implemented a two-year sample cutoff rule for non-recidivists).").

112. The total number of observations in this dataset is 6,216.

113. *See infra* Appendix B, Table 8.

114. *See infra* Appendix B, Table 8b.

115. *See infra* Appendix B, Tables 8, 8b.

116. *See infra* Appendix B, Tables 9, 9b.

117. *See* Bertrand K. Hassani, *Societal Bias Reinforcement Through Machine Learning: A Credit Scoring Perspective*, 1 A.I. & ETHICS 239 (2021); Christophe Hurlin, Christophe

COMPAS serves as a helpful case study for a larger sociotechnical problem that arises when decision-making algorithms are used, as data on mortgages show.[118]

In a previous study, one of us delved into the impact of AI on the mortgage application process and its influence on bias within the mortgage market.[119] That study examined how the implementation of a fair AI model in the mortgage application procedure affects the well-being of various stakeholders.[120] The study used historical mortgage application data made available through the Home Mortgage Disclosure Act (HMDA) in 2019,[121] which included mortgage types, income levels, loan amounts, race, gender, and additional pertinent geographical and personal statistics.[122] This dataset included information about each mortgage application, including the ultimate approval or denial.[123] The paper used a supervised AI model developed to categorize whether each mortgage application should be accepted or rejected.

The paper employed the same techniques as this Article (FACE, FACT, and causal forest) to examine the mortgage loans dataset.[124] It found that the HMDA dataset contains racial bias: Black applicants are less likely to have their mortgage applications approved.[125]

The paper revealed that this racial bias becomes more pronounced in AI prediction results compared to the bias present in the original dataset.[126] In other

---

Pérignon & Sébastien Saurin, *The Fairness of Credit Scoring Models*, MGMT. SCI., Nov. 2024, at 1.

118. *See, e.g.*, Black et al., *supra* note 14, at 62 ("[C]onsider a bank that uses an algorithmic system to make loan decisions. The prediction task can be defined simply: classify individual applicants as creditworthy or not. Because creditworthiness cannot be directly measured, the bank uses some other measure—for example, the likelihood of nonpayment after 6 months—as a proxy." (footnote omitted)).

119. Leying Zou & Warut Khern-am-nuai, *AI and Housing Discrimination: The Case of Mortgage Applications*, 3 A.I. & ETHICS 1271, 1271 (2023).

120. *Id.* at 1271 ("[W]e study how the adoption of a fair AI model in the mortgage application process impacts the welfare of stakeholders.").

121. *Id.*

122. OFF. OF THE COMPTROLLER OF THE CURRENCY, INTERAGENCY CONSUMER LAWS AND REGULATIONS: HOME MORTGAGE DISCLOSURE ACT (HMDA) 23–24 (2021), https://www.occ.treas.gov/publications-and-resources/publications/comptrollers-handbook/files/home-mortgage-disclosure/index-home-mortgage-disclosure.html [https://perma.cc/X2NT-U5UU].

123. The HMDA mandates financial institutions providing mortgage-related products to disclose anonymous information pertaining to mortgage applications and decisions. *See id.* at 1 ("The Home Mortgage Disclosure Act (HMDA) requires certain financial institutions to collect, report, and disclose information about their mortgage lending activity."); 12 U.S.C. § 2803(a).

124. *See* Zou & Khern-am-nuai, *supra* note 119, at 1274.

125. *Id.* at 1271–72*; see also* Andreas Fuster, *Paul Goldsmith-Pinkham, Tarun Ramadorai & Ansgar Walther, Predictably Unequal? The Effects of Machine Learning on Credit Markets*, 77 J. FIN. 5, 7–12 (2022) (examining U.S. mortgage approval data from 2009 to 2013 and finding that Black and Hispanic applicants lose with the incorporation of a machine learning algorithm).

126. Zou & Khern-am-nuai, *supra* note 119, at 1272 ("[W]e . . . find that the AI model not only inherits bias from the training dataset but actually amplifies that bias.").

words, when an AI model is trained based on historical decisions regarding mortgage application approvals and denials, the model captures both the historical approval criteria and the underlying racial bias in the process. When this trained model is deployed without human intervention, the bias in the model is accentuated due to the model's objective of maximizing prediction accuracy. The use of common, readily available AI models to make decisions about whether to approve or decline mortgage applications introduces further racial bias into the prediction process.[127] This bias occurs because the training dataset is historically biased against Black applicants.

The adoption of an AI model exacerbates this existing bias by over 13%.[128] When a commonly used AI model is employed, it incorporates racial bias from the training data without sampling problems and it leads to outcomes that are unfairly skewed against Black individuals.[129] To ensure that the identified bias is not merely a correlation between race and approval decisions, the paper established causal inferences, attributing the disparities in approval decisions to the applicants' race.[130] The AI model, like COMPAS, not only replicated bias from the training dataset but also amplified such bias.[131]

To achieve bias mitigation, the paper employed an algorithm called "Exponentiated Gradient."[132] The selection rate difference, which measures the gap in approval rates between Black applicants and applicants of other groups, was significantly reduced, from 17.40% to 1.46%.[133]

It is important to highlight that the bias reduction led to substantial outcomes. Due to this mitigation, the count of approved Black applicants for mortgages increased by approximately 165%, surging from 4,722 to 12,494 applications.[134] Similarly, applicants of other racial backgrounds also experienced a beneficial impact, increasing the number of approved mortgage applications by approximately 6% (from 148,611 to 157,679 applicants).[135]

This fairness constraint, however, came at the cost of an increased overall false positive rate of the bias-mitigated model, soaring by 362%.[136] Notably, while the

---

127. *Id.* at 1277 ("It is clear that the use of common, off-the-shelf AI models to approve or decline mortgage applicants induces racial bias in the prediction process.").

128. *Id.*

129. *Id.* at 1272 ("With the developed model, we first demonstrate that when a common AI model is used, unsurprisingly, it inherits the ethnicity bias that exists in the training data, producing results that are biased against black applicants.").

130. *Id.* ("[W]e can causally attribute the discrepancy in the approval decision to the ethnicity of the applicants.").

131. *Id.* ("In this exercise, we also find that the AI model not only inherits bias from the training dataset but actually amplifies that bias.").

132. Integrated within the Fairlearn toolkit. *See* BIRD ET AL., *supra* note 106.

133. Zou & Khern-am-nuai, *supra* note 119, at 1280 ("The selection rate difference (i.e., the difference in the approval rate of Black or African American applicants and that of applicants of other races) is reduced from 17.40% to 1.46%.").

134. *Id.* at 1278.

135. *Id.* at 1279 ("[A]pplicants of other races also benefit from bias mitigation as the number of applicants who are approved for the mortgage increases by about 6% from 148,611 applicants to 157,679 applicants.").

136. *Id.* ("Because of this increase, the overall false positive rate of the bias-mitigated model increases by 362%, which may impose a significant burden on financial institutions overall accuracy decreased by only 3%, prediction accuracy for Black applicants decreased by 15% (accuracy for non-Black applicants dropped by only 2%).[137] Although the overall prediction accuracy diminished with this bias mitigation technique, the volume of mortgage applicants notably grew by 165% for Black applicants and by 6% for applicants of other groups.[138] Given the higher number of mortgages approved, there was an average decrease of 36% in the loan amounts granted.[139]

The elevated false positive rate could concern financial institutions adopting the fair AI model. Unlike the case of COMPAS, where the cost of fairness argument posits that it is externalized to society, in this case the "cost of fairness" would be paid by the decision-maker (the mortgage lender), since false positive predictions for this type of prediction task can be costly to a financial institution.[140] But these concerns should be mitigated if the model's outcome variable "similarity with people who successfully paid loans in the past" is racially skewed like COMPAS's outcome variable is.[141]

Two real-world implications of these findings match the implications of this Article's study. First, the direct use of a common AI model to approve mortgage applications without human intervention, as with risk assessment in criminal justice, should be discouraged since it amplifies historical bias from training datasets. Second, when a fairness-adjusted AI model is used to mitigate such a bias, decision-makers (in this case, mortgage lenders) might be uneasy about whether and to what extent the fairness-adjusted model is paid with a cost in accuracy. They should not be.

## IV. IMPLICATIONS FOR FAIR AI DECISION-MAKING

### *A. Who Pays the Cost of Fairness?*

Despite the racial bias revealed in the COMPAS dataset, many believe that reducing discrimination will come at the cost of accuracy.[142] An increment in false positives, this argument goes, represents the "cost of fairness" incurred by society,

---

that adopt the fair AI model.").

137. *Id.* at 1278–79 ("[I]t is worth noting that even though the overall accuracy decreases by about 3 percentage points, the prediction accuracy for Black or African American applicants decreases by almost 15 percentage points while the prediction accuracy for applicants of other races decreases by almost 2 percentage points only.").

138. *Id.* ("Meanwhile, the number of mortgage applicants increases by 165% for Black or African applicants and 6% for applicants of other races.").

139. *Id.* at 1280 ("[E]ven though more Black or African American applicants receive approval decisions for their mortgage applications, the average loan amount significantly decreases by 36%.").

140. *Cf. id.* at 1272 ("[W]hen the fair AI model is used to mitigate such a bias, it is the mortgage lender who pays for the cost of fairness.").

141. *Cf. id.* at 1280 ("Additional considerations from multiple perspectives, including consumer surplus, social welfare, and legal issues, are still lacking in the literature, as well as empirical results from the real-world implementations of fair AI models.").

142. *See, e.g.*, Corbett-Davies et al., *Algorithmic Decision Making and the Cost of Fairness*, *supra* note 12.

where the "cost of fairness" refers to an increased number of future criminals who are released when a fair AI model is used. This argument extends beyond policy circles into academia and the popular press.[143] The policy corollary to this argument posits that society should not "pay" for fairness by having more dangerous individuals released.[144]

The first problem with this argument is demonstrated by the concept of model multiplicity, put forth by computer science scholars, which suggests that "there are almost always multiple possible models with equivalent accuracy for a given prediction problem . . . ."[145] Emily Black and others explain that "when an algorithmic system displays a disparate impact, model multiplicity suggests that other models exist that perform equally well but have less discriminatory effects."[146] In other words, when a model such as COMPAS has disparate effects, there is almost always a less discriminatory algorithm that could serve as an alternative without entailing a loss of performance.[147] According to Black et al., it should be the responsibility of developers and deployers to test for disparate impact and find these alternative algorithms.[148]

The second problem with this argument is that the COMPAS training dataset is racially biased because arrest data in the United States are racially biased.[149] COMPAS is trained on arrest data for crimes across the United States.[150] The company's practitioner's guide defines recidivism as "new offenses subsequent to the COMPAS assessment date," where an offense is "a new misdemeanor or felony offense within two years of the COMPAS administration date."[151] ProPublica, on the other hand, defined recidivism in its study as "a finger-printable arrest involving a charge and a filing for any uniform crime reporting (UCR) code."[152] This matters because, as the next Section details, people with a "high-risk" outcome variable and "future criminals" are not the same thing.[153]

---

143. *See, e.g.*, *id.* at 800 (arguing that algorithms not subject to fairness constraints maximize public safety); Han Zhao & Geoffrey J. Gordon, *Inherent Tradeoffs in Learning Fair Representations*, 23. J. MACH. LEARNING RSCH., Jan. 2022, at 1, 2 (presenting a cost for statistical parity in terms of algorithmic accuracy); Irene Y. Chen, Fredrik D. Johansson & David Sontag, *Why Is My Classifier Discriminatory?*, 33RD CONF. ON NEURAL INFO. PROCESSING SYST. 3539, 3539 (2018) (arguing that balancing accuracy with fairness leads to undesirable outcomes and that unfairness should be addressed at data collection, not by constraining the model); Irineo Cabreros, Opinion, *Why an Algorithm Can Never Truly Be 'Fair'*, L.A. TIMES (Mar. 27, 2022, 3:10 AM), https://www.latimes.com/opinion/story/2022-03-27/algorithms-unfair-racial-bias-math [https://perma.cc/5GE3-SGGD] ("[A]lgorithms can attain higher accuracy if they aren't also required to perform equitably. . . . [E]verything is more difficult when constraints are added. . . . The most accurate algorithm is probably more accurate than the most accurate *equitable* algorithm." (emphasis in original)).

144. *See, e.g.*, Chohlas-Wood et al., *Learning to Be Fair*, *supra* note 13; Corbett-Davies et al., *Algorithmic Decision Making and the Cost of Fairness*, *supra* note 12, at 802–03; Corbett-Davies et al., *The Measure and Mismeasure of Fairness*, *supra* note 11; Chohlas-Wood et al., *Designing Equitable Algorithms*, *supra* note 13.

145. Black et al., *supra* note 14.

146. *Id.* at 57.

147. *See id.*

148. *Id.* at 85.

149. Ngozi Okidegbe, *Discredited Data*, 107 CORNELL L. REV. 2007, 2007 (2022).

150. That is, arrests for misdemeanors do not count, and arrests for nonviolent crimes do count. *See* Larson et al., *supra* note 5; Angwin & Larson, *supra* note 26; *see also Crime in the United States: Violent Crime*, FED. BUREAU OF INVESTIGATION (2010), https://ucr.fbi.gov/crime-in-the-u.s/2010/crime-in-the-u.s.-2010/violent-crime [https://perma.cc/U7KK-42WK] (classifying crimes between violent and nonviolent).

While the average bias in arrest data in the United States is difficult to calculate (a possible estimation may be our analysis of COMPAS data in this Article), there is abundant empirical evidence of arrests in the United States being racially biased.[154] One meta-analysis, examining quantitative research estimating the effects of race on the police decision to arrest, shows with strong consistency that Black suspects are more likely to be arrested than white suspects.[155] Some criminologists estimate that police are more than twice as likely, on average, to arrest Black individuals than white individuals for an offense.[156] For example, according to National Survey on Drug Use and Health data, Black individuals constitute 29% of arrests for drug offenses while they are estimated to constitute only 12.5% of illegal drug users.[157] Black individuals and white individuals are equally likely to consume marijuana illegally, but Black individuals are arrested at approximately twice the rate of white individuals for marijuana possession.[158] Some estimates indicate that Black individuals are 235% more likely than white individuals to be arrested for drug-related crimes while being less likely to engage in those crimes.[159]

The disparity goes beyond drug-related arrests.[160] Empirical data controlling for effects such as socioeconomic conditions, neighborhood crime rates, demographic turnover, and policing strategies, do not support the idea that Black individuals commit more violent crimes than white individuals, even though Black individuals are arrested at a higher rate.[161] FBI data indicate that Black individuals are arrested for violent crimes at a much higher rate than white individuals[162] but get convicted at a lower rate.[163] A similar gap has been found in motor vehicle police searches.[164] Another study, examining New York's stop and frisk policy, found an arrest rate of 2.5% for Black individuals and 11% for white individuals—meaning that when black individuals were stopped, it was "typically on the basis of less evidence."[165] Older studies find smaller but still significant disparities.[166] This disparity, often called a "hit rate" disparity, indicates that the higher rate of arrest is racially biased.[167]

---

151. NORTHPOINTE, PRACTITIONER'S GUIDE TO COMPAS CORE 27 (2015).

152. Larson et al., *supra* note 5 (quoting Tim Brennan, William Dieterich & Beate Ehret, *Evaluating the Predictive Validity of the Compas Risk and Needs Assessment System*, 36 CRIM. JUST. & BEHAV. 21, 26 (2009)).

153. *See infra* Section IV.B.

154. *See, e.g.*, Yu Du, *Racial Bias Still Exists in Criminal Justice System? A Review of Recent Empirical Research*, 37 TOURO L. REV. 79, 85–91 (2021).

155. Tammy Rinehart Kochel, David B. Wilson & Stephen D. Mastrofski, *Effect of Suspect Race on Officers' Arrest Decisions*, 49 CRIMINOLOGY 473, 490–91 (2011) (graphically depicting that minorities and Black suspects have a higher probability of arrest across data sets).

156. Brendan Lantz & Marin R. Wenger, *The Co-Offender as Counterfactual: A Quasi-Experimental Within-Partnership Approach to the Examination of the Relationship Between Race and Arrest*, 16 J. EXPERIMENTAL CRIMINOLOGY 183, 199 (2020).

157. Du, *supra* note 154, at 83–84.

158. *ACLU Report: Racial Disparities Persist in Marijuana Possession Arrests*, NORML (Apr. 23, 2020), https://norml.org/news/2020/04/23/aclu-report-racial-disparities-persist-in-marijuana-possession-arrests [https://perma.cc/YJ7Z-N8LA].

159. Ojmarrh Mitchell & Michael S. Caudy, *Examining Racial Disparities in Drug Arrests*, 32 JUST. Q. 288, 307–09 (2015) (comparing rates of drug arrests for Black and white individuals at age 27).

160. *See, e.g.*, Du, *supra* note 154, at 88–89; Rinehart Kochel et al., *supra* note 155, at 473.

Note that the data surveyed above comparing arrest numbers with conviction numbers indicate that racial bias exists in arrest, even assuming that convictions are not racially biased. Given the evidence that convictions may also contain racial bias,[168] the bias in arrest may be even larger. All in all, the data show that arrest is a biased, and therefore flawed, metric for recidivism.

---

161. Gregory DeAngelo, R Kaj Gittings & Anita Alves Pena, *Interracial Face-to-Face Crimes and the Socioeconomics of Neighborhoods: Evidence from Policing Records*, 56 INT'L REV. L. & ECON. 1, 5–6 (2018) (analyzing data on homicide, robbery, assault, and use of weapons); *see* Daniel P. Mears, Joshua C. Cochran & Andrea M. Lindsey, *Offending and Racial and Ethnic Disparities in Criminal Justice: A Conceptual Framework for Guiding Theory and Research and Informing Policy*, 32 J. CONTEMP. CRIM. JUST. 78, 80 (2016) (estimating Black individuals to be arrested at over twice the rate of white individuals); THE SENT'G PROJECT, REPORT TO THE UNITED NATIONS ON RACIAL DISPARITIES IN THE U.S. CRIMINAL JUSTICE SYSTEM 2 (2018) (showing that Black individuals form approximately twice the percentage of arrests than they do the percentage of the population); Cydney Schleiden, Kristy L. Soloski, Kaitlyn Milstead & Abby Rhynehart, *Racial Disparities in Arrests: A Race Specific Model Explaining Arrest Rates Across Black and White Young Adults*, 37 CHILD & ADOLESCENT SOC. WORK J. 1, 1 (2020) (estimating a sevenfold difference in each Black individual's likelihood of arrest compared to white individuals).

162. *2015 Crime in the United States*, FED. BUREAU INVESTIGATION (2015), https://ucr.fbi.gov/crime-in-the-u.s/2015/crime-in-the-u.s.-2015/tables/table-43/#overview [https://perma.cc/D6WZ-YSR8]; *see also* Emma Pierson et al., *A Large-Scale Analysis of Racial Disparities in Police Stops Across the United States*, 4 NATURE HUM. BEHAV. 736, 736 (2020) (analyzing cross-U.S. data of about 100 million traffic stops).

163. BUREAU OF JUST. STATS., FELONY SENTENCES IN STATE COURTS, 2006 - STATISTICAL TABLES 17 tbl.3.2 (2010) (National Judicial Reporting Program indicating that 39% of those convicted of violent crimes are Black and 58% are white).

164. *See* John Knowles, Nicola Persico & Petra Todd, *Racial Bias in Motor Vehicle Searches: Theory and Evidence*, 109 J. POL. ECON. 203, 204–06 (2001).

165. Sharad Goel, Justin M. Rao & Ravi Shroff, *Precinct or Prejudice? Understanding Racial Disparities in New York City's Stop-and-Frisk Policy*, 10 ANNALS APPLIED STAT. 365, 375–82 (2016); *see also* Pierson et al., *supra* note 162.

166. *See, e.g.*, Andrew Gelman, Jeffrey Fagan & Alex Kiss, *An Analysis of the New York City Police Department's "Stop-and-Frisk" Policy in the Context of Claims of Racial Bias*, 102 J. AM. STAT. ASS'N 813, 820 (2007); Eric A. Stewart, Eric P. Baumer, Rod K. Brunson & Ronald L. Simons, *Neighborhood Racial Context and Perceptions of Police-Based Racial Discrimination Among Black Youth*, 47 CRIMINOLOGY 847, 865 (2009) (showing the impact of location in unjustified stop and frisk).

167. Decio Coviello & Nicola Persico, *An Economic Analysis of Black-White Disparities in the New York Police Department's Stop-and-Frisk Program*, 44 J. LEGAL STUD. 315, 338 (2015); Du, *supra* note 154, at 87.

It is worth noting that the bias observed in the COMPAS dataset in our study is similar to the bias in arrests observed and quantified in prior works that use other variables to estimate criminality. Our estimates of bias, in other words, support prior qualitative claims.[169] In our study, the racial bias in the COMPAS dataset, defined as the gap between the recidivism rate of Black and white individuals who are otherwise similar, ranges from 12.42% to 19.01%, depending on the identification method.[170] A recent study estimating the effect of offenders' race on the probability of being held during pretrial,[171] similarly, found that the racial gap for the hold rates between Black and white individuals is approximately 11%.[172] While this equivalence is not conclusive proof that the COMPAS dataset bias that we find (prior to being amplified by the algorithm) is driven by racial bias in arrest data, it is a strong indication that it is likely to be.

### *B. The Impact of Human-Chosen Metrics*

A recidivism algorithm's outcome variable is inevitably chosen by its designers as an estimate of an ideal decision criterion. In the case of COMPAS, decision-makers in theory base parole or bail decisions on the risk that the defendant will commit a violent crime during the period when they would otherwise have remained in prison. The algorithm uses the outcome variable "likelihood of rearrest" to estimate the risk that the defendant will commit a violent crime because there are no data on reoffense. As this substitution may be inevitable, it is important to recognize when it is inaccurate or introduces bias.[173] Predicting an arrest is analytically and practically distinct from predicting a conviction—and predicting any misdemeanor or felony is analytically and practically distinct from predicting a violent crime.[174]

In AI systems, one is often interested in unobservable concepts such as "risk to society."[175] Many discriminatory outcomes of AI systems result from an overlooked mismatch between those concepts and their operationalization.[176] Recidivism refers to the relapse into criminal behavior after a prior conviction, while rearrest signifies a new arrest after release.[177] As Abigail Jacobs and Hanna Wallach explain, "By assuming that arrests are a reasonable proxy for crimes committed, COMPAS fails to account for false arrests or crimes that do not result in arrests. . . . [N]o computational system can ever wholly and fully capture the substantive nature of crime by using arrest data as a proxy."[178]

---

168. *See* Du, *supra* note 154, at 91–98.

169. *See* Okidegbe, *supra* note 149.

170. *See infra* Appendix B, Table 3.

171. Frank McIntyre & Shima Baradaran, *Race, Prediction, and Pretrial Detention*, 10 J. EMPIRICAL LEGAL STUD. 741, 741 (2013) (using data from nationally representative State Court Processing Statistics on felony defendants); *see also* Pierson et al., *supra* note 162.

172. McIntyre & Baradaran, supra note 171, at 753.

173. *See* Rachel L. Thomas & David Uminsky, *Reliance on Metrics Is a Fundamental Challenge for AI*, PATTERNS, May 13, 2022, at 1, 1–4.

174. Washington, *supra* note 98, at 143 ("Algorithms, like COMPAS, operationalize risk of recidivism as predicting those who are likely to be arrested again, which rarely considers geographic structural conditions. Predicting a misdemeanor arrest or felony arrest is analytically different from predicting misdemeanor conviction or felony conviction." (footnotes omitted)).

175. *See* Abigail Z. Jacobs & Hanna Wallach, *Measurement and Fairness*, FACCT '21: PROCS. 2021 ACM CONF. ON FAIRNESS, ACCOUNTABILITY & TRANSPARENCY 375.

176. *Id.* at 375.

While recidivism algorithms are meant to assess whether a defendant is likely to commit a violent crime if released on parole,[179] whether the defendant will commit such an offense if released is unknowable at the time of the decision.[180] Even if it were a perfectly accurate estimate of violent crime convictions, the algorithm's output (i.e., likelihood of rearrest within two years) may under- or overestimate the actual likelihood that a particular prisoner will *recidivate*. The use of biased training data explains this phenomenon—as Black individuals are rearrested more often than white individuals.[181]

A mismatch between the outcome variable and the ideal grounds for a decision can be particularly significant in machine-learning-based decision-making algorithms because the machine learning process constrains the selection of outcome variables. To train a model that computes reasonably appropriate predictions for an outcome variable, the algorithm's designers must have access to a sufficiently large set of data correlating feature values to outcome values. Such datasets are ordinarily available for only a limited selection of outcome variables. The outcome variables for which such data are available may not be close substitutes for the decisions' ideal grounds. As a result, designers may have to prioritize either the accuracy with which their machine learning model works *for a given outcome variable* or with the reliability of that outcome variable *for the ideal bases for decisions*.[182]

Our analysis of COMPAS data shows, in line with prior findings, that bias exists in the dataset: Black individuals received higher risk scores and higher error rates. Whether this phenomenon is a sample problem depends on whether increasing the sample size would remedy the problem. Interpreting these results using the literature on the topic suggests that it would not: Bias comes from incorporating the racial bias of rearrest data as compared to actual recidivism.[183] The dataset used to train the model is biased because of historical bias in the criminal justice system before the use of AI.[184] Our results show that a standard AI model amplifies that racial bias by 10.66 percentage points (a 20.58% increase).

---

177. James L. Johnson, *Comparison of Recidivism Studies: AOUSC, USSC, and BJS*, 81 FED. PROB. 52, 52 (2017) ("Recidivism is commonly defined as reengaging in criminal behavior after receiving a sanction or undergoing an intervention for a previous crime.").

178. *Id.* at 380 (footnote omitted).

179. Jessica M. Eaglin, *Constructing Recidivism Risk*, 67 EMORY L.J. 59, 75 (2017) ("To glean information from that base population . . . requires developers to translate a problem—here, recidivism—into a formal question about variables. . . . Developers frame this question around what they would like to know at sentencing: whether this person will commit a crime in the future." (footnotes omitted)).

180. *See, e.g.*, Dressel & Farid, *supra* note 17.

181. Julia Angwin, Jeff Larson, Surya Mattu & Lauren Kirchner, *Machine Bias*, PROPUBLICA (May 23, 2016), https://www.propublica.org/article/machine-bias-risk-assessments-in-criminal-sentencing [https://perma.cc/Z69B-HRP3].

182. *See* John Nay & Katherine J. Strandburg, *Generalizability: Machine Learning and Humans-in-the-Loop*, *in* RESEARCH HANDBOOK ON BIG DATA LAW 285, 288–89 (Roland Vogl ed., 2020); Deborah Hellman, *Measuring Algorithmic Fairness*, 106 VA. L. REV. 811, 841 (2020) ("Some traits simply cannot be measured directly, and proxies will be the best we can do.").

COMPAS is not biased as a measure of rearrest—it is biased as a measure of reoffense.[185] By using prediction of rearrest as a proxy for reoffense,[186] COMPAS incorporates social biases that distort the relationship between offending and being arrested.[187] So, COMPAS is biased because social biases interfere with the relationship between historical arrests and actual recidivism.[188] In other words, the problem is the human decision to use rearrest as a proxy for reoffending.

This substitution faces two problems. First, as rearrests for violent offenses are relatively rare, it is difficult to train a model to predict them because an appropriate sample of data for that outcome variable may be unavailable. So designers must choose between the less plentiful, but more meaningful, data of arrests for violent recidivism and the more plentiful, but less meaningful, data of arrests for any offense.[189] For example, an individual who is unlikely to commit a violent crime if released pending trial might be highly likely to be rearrested for committing a less serious crime, such as possessing drugs, but using these data muddles what the decision-maker is trying to predict.

Second, the relevant outcome to be measured is the likelihood of *committing* a (violent) crime, and arrest is an imperfect estimate for it.[190] Data about rearrests are biased by the fact that only those offenders and nonoffenders who are apprehended are counted.[191] For example, an individual might be more likely than others to be

---

183. Larson et al, *supra* note 5; Angwin et al., *supra* note 181.

184. *See* Dressel & Farid, *supra* note 17.

185. *See* Corbett-Davies et al., *Algorithmic Decision Making and the Cost of Fairness*, *supra* note 12, at 803 fig.2 (noting that calibration of COMPAS was satisfactory in the sense that the predictive accuracy of rearrest was the same for both groups).

186. *See* Stevenson, *supra* note 26, at 303.

187. *See* Michael Zanger-Tishler, Julian Nyarko & Sharad Goel, *Risk Scores, Label Bias, and Everything but the Kitchen Sink*, 10 SCI. ADVANCES, Mar. 29, 2024, at 1, 2 ("[A]rrests and convictions are not direct measures of public safety risks. Instead, they merely act as proxies, making these risk assessment tools susceptible to label bias."); *see also* Pauline T. Kim, *Auditing Algorithms for Discrimination*, 166 U. PA. L. REV. ONLINE 189, 191 (2017) [hereinafter Kim, *Auditing Algorithms for Discrimination*] (explaining that bias in machine learning can be caused by social processes that are reflected in the data).

188. *See* Kim, *Auditing Algorithms for Discrimination supra* note 187, at 191; Kim, *Race-Aware Algorithms*, *supra* note 39, at 154.

189. *See* Kim, *Race-Aware Algorithms*, *supra* note 39, at 1544 ("The designers must make difficult choices each step of the way, involving . . . subjective judgments and the weighing of values. Each of these choices can be consequential in shaping the final model and the results it produces.").

190. *See* Zanger-Tishler et al., *supra* note 187, at 4–5 ("[I]t is difficult—and perhaps impossible—to directly estimate the risk of true offending. This is partly because criminal behavior that is not reported to the police will not be included in administrative records." (footnote omitted)).

191. *See* Skeem & Lowenkamp, *supra* note 25, at 700; *see also* Latanya Sweeney, *Discrimination in Online Ad Delivery*, 11 ACM QUEUE, Mar. 1, 2013, at 1, 10 (showing that arrested because of characteristics such as where they live or their social group, even if they are equally as likely as others to commit a crime.[192] They may even be more likely to be mistakenly detained simply because another machine learning algorithm used by the local police, such as PredPol or HunchLab, employs an inaccurate outcome variable for the likelihood of violent crime as well.[193] The algorithms' output (i.e., likelihood of a new arrest for a misdemeanor or felony within two years[194]) can under- or over-estimate the actual likelihood that any particular defendant will commit a (violent) crime if released.

COMPAS is not the only algorithm to have a disparate impact when making risk assessment predictions. Risk-assessment algorithms beyond COMPAS also combine information about an individual's prior arrests, prior employment, gang affiliation, and behavior in prison.[195] The Chicago Police Department's Strategic Subject List is

---

arrest records advertisements are more likely to appear on searches for Black-sounding names).

192. Zanger-Tishler et al., *supra* note 187, at 3 ("[W]e would accordingly expect the individual in the heavily policed area to be less likely to engage in future criminal behavior. Thus, using information about one's neighborhood to predict future arrests (the proxy label) correctly tells us that the individual living in the heavily policed neighborhood is more likely to be rearrested, but it incorrectly suggests that individual is also more likely to engage in future criminal behavior (the true label). So, when predicting arrests as a proxy for behavior, it is better in this case to exclude information on one's neighborhood.").

193. Aaron Shapiro, *Reform Predictive Policing*, 541 NATURE 458, 458 (2017) ("Geospatial modeling generates risk profiles for locations. . . . [A]lgorithms that have been trained using crime and environmental data predict where and when officers should patrol to detect or deter crime."); *see New Model Police*, THE ECONOMIST (June 7, 2007), https://www.economist.com/united-states/2007/06/07/new-model-police [https://perma.cc/H3FP-9T3K].

194. NORTHPOINTE, *supra* note 151, at 27 ("The recidivism risk scale was developed to predict new offenses subsequent to the COMPAS assessment date. The outcome used for the original scale construction was a new misdemeanor or felony offense within two years of the COMPAS administration date."); Larson et al., *supra* note 5 ("For most of our analysis, we defined recidivism as a new arrest within two years. We based this decision on Northpointe's practitioners guide.").

195. Vincent M. Southerland, *The Intersection of Race and Algorithmic Tools in the Criminal Legal System*, 80 MD. L. REV. 487, 515 (2021) ("[New Jersey's pretrial risk assessment's] factors assessed amount to the accused's age at current arrest, criminal history . . . prior failures to appear, and prior carceral sentences."); John Logan Koepke & David G. Robinson, *Danger Ahead: Risk Assessment and the Future of Bail Reform*, 93 WASH. L. REV. 1725 (2018); Spencer G. Lawson, Staci J. Rising, Eric Grommon & Evan M. Lowder, *Pretrial Risk Assessment Tool Adoption and Pretrial Operations*, *in* HANDBOOK ON PRETRIAL JUSTICE 276–95 (Christine S. Scott-Hayward, Jennifer E Copp & Stephen Demuth eds., 2022); *see, e.g.*, *About the Public Safety Assessment*, ADVANCING PRETRIAL POL'Y & RSCH., https://advancingpretrial.org/psa/about/ [https://perma.cc/4KAU-9SF4]; *see also* SCAN OF PRETRIAL PRACTICES, PRETRIAL JUST. INST. 27 (2019), https://www.pretrial.org/files/resources/scanofpretrialpractices.pdf [https://perma.cc/9QML-L9YD] (explaining widespread use of the Public Safety Assessment); ADVANCING PRETRIAL POL'Y & RSCH., PUBLIC SAFETY ASSESSMENT: HOW IT WORKS (2020), https://cdn.filestackcontent.com/security=policy:eyJleHBpcnkiOjQwNzg3NjQwMDAsImN hbGwiOlsicGljayIsInJlYWQiLCJ3cml0ZSIsIndyaXRlVXJsIiwic3RvcmUiLCJjb252ZXJ0Ii

another example of the same problem.[196] The Strategic Subject List's algorithm predicts the likelihood of a person being involved in gun violence in the future, either as a victim or perpetrator.[197] However, because the list was allegedly used by the police department as an informal suspect list for crimes involving gun violence, it was shown to be predictive not of involvement in future gun violence but of the probability of being arrested in the future.[198] Similarly, another risk assessment algorithm used at the federal level for probation was found to assign a higher average score of (post-conviction) risk assessment to Black individuals.[199] A study concluded that bias in this algorithm was unlikely because 66% of the racial difference was attributable to criminal history and, according to the study, because criminal history is not a proxy for race.[200] But, as we argued, criminal history does affect the relationship between race and future arrest.

Several factors that influence the criminal justice system and individuals' pathways following release make the correlation between rearrest and recidivism imperfect. One is the initial arrest and conviction discrepancy. Not all arrests result in convictions, and not all convictions lead to rearrests.[201] Legal proceedings, evidence availability, and prosecutorial discretion can affect this disparity.[202] A second is diversity of offenses: Individuals can be rearrested for a different, sometimes minor, offense, not necessarily equivalent to their initial conviction. A third is release conditions: Parole, probation, and other release conditions influence rearrest rates more directly than recidivism rates.[203] Finally, systemic biases: Racial and socioeconomic disparities affect the likelihood of rearrest and recidivism independently.

The literature has shown that the process of using a proxy "necessarily involves making assumptions."[204] And this reality includes the choice of outcome variables.

---

wicmVtb3ZlIiwicnVuV29ya2Zsb3ciXX0=,signature:9df63ee50143fbd862145c8fb4ed2fcc17d068183103740b1212c4c9bc858f63/5gCeQzRTuWKKCf5WL7mg [https://perma.cc/9QXS-3CTC] (explaining data sources and outcome variables).

196. *See* Ashley S. Deeks, *Predicting Enemies*, 104 VA. L. REV. 1529, 1543–45 (2018).

197. *Id.* at 1543.

198. *See* Jessica Saunders, Priscillia Hunt & John S. Hollywood, *Predictions Put into Practice: A Quasi-Experimental Evaluation of Chicago's Predictive Policing Pilot*, 12 J. EXPERIMENTAL CRIMINOLOGY 347, 363–64 (2016).

199. *See* Skeem & Lowenkamp, *supra* note 25, at 685.

200. *Id.* at 698.

201. John P. Walters & David Tell, *Criminal Justice Reform and the First Step Act's Recidivism Reduction Provisions: Preliminary Issues for Policymakers*, HUDSON INST. (Jan. 18, 2019), https://www.hudson.org/domestic-policy/criminal-justice-reform-and-the-first-step-act-s-recidivism-reduction-provisions-preliminary-issues-for-policymakers [https://perma.cc/CK2T-QGLU] ("Not all rearrests result in reconviction, of course, and not all reconvictions result in reincarceration.").

202. THE NAT'L ACADS. OF SCIS., ENG'G, & MED., THE LIMITS OF RECIDIVISM: MEASURING SUCCESS AFTER PRISON 45 (Richard Rosenfeld & Amanda Grigg eds., 2022) ("The determination of the convicted offenses reflects decisions of prosecutors, defense attorneys, and judges or juries, and the records of offense behaviors are based on criminal statutes, not offense-specific behaviors.").

203. *Id.* at 19–20.

204. Jacobs & Wallach, *supra* note 175, at 375.

Acknowledging so helps explain why COMPAS's use of "likelihood of rearrest" to measure "risk of reoffense" is flawed, thereby amplifying racial biases in arrest data. This explains much of the critique in the literature, which indicates that implementing risk assessment tools does not automatically translate to the desired outcomes.[205] Megan Stevenson, for example, cites the passage of HB 463 in Kentucky, which "made use of pretrial risk assessment mandatory" yet instead led to an increase in failure-to-appear rates and pretrial crime.[206] Kentucky's HB 463 illustrates how outcome variable selection can interfere with the ability of algorithmic risk assessments, like COMPAS, to deliver on their promises of neutrality. Findings of regional disparities in pretrial detention rates (i.e., rural areas saw a greater decrease in detention compared to urban areas after HB 463) highlight how many factors (e.g., access to bail money and the charged offense) interact with risk assessment tools in unpredictable ways precisely because of this choice of outcome variables.[207]

Due to this bias, releasing people who have a high likelihood of being arrested in the future is not necessarily a socially costly outcome. Because "rearrest" is meant to gauge recidivism, reducing the accuracy of the model for that variable may not reduce accuracy toward what the model is supposed to predict.[208] In other words, these algorithms would have lower predictive accuracy with regard to their human-determined outcome variable (i.e., lower ability to identify similarity with other individuals who have been rearrested), but not necessarily with regard to the target. More importantly, if one believes that COMPAS was biased for the target (likelihood of recidivism), one should also believe that, after adding a fairness constraint, COMPAS should not have lower predictive accuracy for the desired prediction. The adjustment would correct for the error between the outcome variable and the target, i.e., between the likelihood of rearrest and the likelihood of reoffense.

An analysis of judges' sentencing practices in response to the use of risk assessments revealed that judges in Virginia initially changed their sentencing practices by using the risk assessment tool but later deviated, becoming less reliant on it.[209] That study argues that strict adherence to the algorithm's recommendations would have led to both benefits and costs—benefits being a "sharp decrease in . . . the probability of incarceration" and costs being "a small increase in recidivism" and an "increase in relative sentence length for young defendants."[210] More than this costs and benefits analysis, however, the study provides support for the inherent limitations of using proxies in risk assessment, specifically that proxies can complicate the pursuit of multiple objectives in sentencing.[211] Judges, even when presented with an algorithmic risk assessment, considered factors beyond the risk of reoffending, such as the age, gender, and race of the defendant.[212] Notably, the authors argue "[a]n algorithm designed to reduce bias in the scores will not necessarily reduce bias in use."[213] Indeed, the impact of any algorithm is shaped by the human decision-makers who use it, and the deployment of an algorithm to streamline decision-making processes is not a silver bullet.

---

205. *See* Stevenson, *supra* note 26.

206. *Id.* at 308.

207. *Id.* at 364 ("[T]here are regional disparities in the likelihood of being detained pretrial. . . . Before HB 463, rural defendants were about 8 percentage point[s] more likely to be detained pretrial than those living in cities or suburban areas. However, this gap shrunk and then reversed itself over time. The gap shrunk partly because rural regions responded more to HB 463 than non-rural regions. It also shrunk because the release rate dropped precipitously for non-rural regions over the six years of analysis: from a high of about 70% in January 2010 to a low of 55% in January 2016." (footnote omitted)).

208. *See, e.g.*, Michael Feldman, Sorelle A. Friedler, John Moeller, Carlos Scheidegger & Suresh Venkatasubramanian, *Certifying and Removing Disparate Impact*, *in* PROCS. 21TH ACM SIGKDD INT'L CONF. ON KNOWLEDGE DISCOVERY & DATA MINING 259 (2015) (developing a method to remove information on protected categories without reducing accuracy by preserving each individual's rank orthogonal to their class membership and showing that several preprocessing techniques eliminate discrimination with a minimal loss in accuracy).

209. Megan T. Stevenson & Jennifer L. Doleac, *Algorithmic Risk Assessment in the Hands of Humans*, 16 AM. ECON. J: ECON. POL'Y, 382, 382–83 (2024).

In sum, choosing to engage in a prediction, choosing an outcome variable, and choosing a measure for algorithmic accuracy are normative exercises. But, once those three choices are made, the discrepancy between the outcome variable in an algorithm like COMPAS and what the system aims to predict is not a normative point; it is a factual one. The policy concern as to the cost of fairness is thus overblown.

### *C. The Cost of AI Discrimination*

The reason why a racial disparity occurs matters for determining who pays a cost for repairing it—which, in turn, has implications intersecting with antidiscrimination law. The concept of statistical discrimination illustrates why a model trained with representative data can lead to disparate impact discrimination, as we argue COMPAS does.[214]

Imagine an employer-deployed algorithm that cannot observe each worker's skill level. The algorithm, however, captures two things. First, it captures education level, which it uses as a signal about each person's skill. Second, it captures each person's group identity—whether they belong to group *A* or *B*. Determining whether a disparate impact exists involves determining why two equivalent workers from different groups might be treated differently with representative data.[215] Specifically, stereotyping is a type of differential treatment under statistical discrimination that takes place when skill signals are equally informative for members of both groups, but group *A* has, on average, fewer education credentials. The algorithm might take group membership to be informative and estimate an employee from group *A* to have lower *expected* skill than an employee from group *B* with the same signals. Therefore, they may receive a lower salary or be offered fewer jobs.

---

210. *Id.* at 383–84.

211. *See id.* at 411 ("Why didn't risk assessment use lead to a detectable decrease in crime? We expect that this is partly because conflicting objectives reduced the potential gains from risk assessment.").

212. *See id.* at 404 ("Several things stand out from this analysis. First, judges give young defendants sentences that are almost 30 percent shorter than those given to older defendants at the same risk level. (Young is defined as being under the age of 23.) Second, judges give Black defendants sentences that are 12–24 percent longer than White defendants with the same risk score. Finally, defendants who are female, unemployed, or married receive favorable treatment (4–18 percent shorter sentences). These patterns hold even when controls for the guidelines-recommended sentence and judge are included.").

213. *Id.* at 405.

214. *See* Edmund S. Phelps, *The Statistical Theory of Racism and Sexism*, 62 AM. ECON. REV. 659, 661 (1972); Kenneth Arrow, *The Theory of Discrimination*, *in* DISCRIMINATION IN LABOR MARKETS 3 (Orley Ashenfelter & Albert Rees eds., 1973).

Recidivism from COMPAS-examined individuals is unobservable like *A*'s and *B*'s levels of skill is in the hypothetical. Like *A* individuals were disadvantaged by their group members having fewer education credentials on average, which the algorithm used as a signal of skill, Black individuals under COMPAS are disadvantaged by their group members having higher arrest rates on average.[216] While the COMPAS *model* does not consider race, the COMPAS *dataset* does so indirectly by its choice of outcome variable. In the COMPAS dataset, the choice of using rearrest is what generates a statistical bias, i.e., a difference between the estimator's expected value and its true value, as in the hypothetical.

The disparate impact doctrine in antidiscrimination law addresses statistical discrimination. This doctrine focuses on practices that, while appearing neutral, disproportionately affect certain groups.[217] In employment, housing, and lending, if a policy leads to statistical disparities among protected classes, it can be deemed discriminatory unless a valid business necessity exists.[218] In lending, for instance, if a credit algorithm results in higher loan rejections for a particular group, it might be considered discriminatory even without explicit bias.[219]

Because disparate impact discrimination is a problem of adversely affecting protected populations without a classification bias, it can be applied to this data problem.[220] Just as discrimination can exist when hiring policies unintentionally

---

215. *See* Shelly Lundberg & Richard Startz, *On the Persistence of Racial Inequality*, 16 J. LAB. ECON. 292, 292–95 (1998) (introducing a model showing that statistical discrimination in competitive markets and without differences in average human capital introduces inefficiencies in the system; due to statistical discrimination, minorities face lower incentives to invest in human capital, community social capital is lowered, and they develop lower levels of productivity).

216. *See supra* Section IV.A.

217. Black et al., *supra* note 14, at 116 ("Civil rights laws like Title VII, the FHA, and the ECOA prohibit not only disparate treatment (commonly described as intentional discrimination) but also disparate impact (facially neutral practices that have unfairly disparate effects on disadvantaged groups).").

218. *Id.* at 72–74 (discussion of the disparate impact framework).

219. *Id.* at 65–66 ("Apart from who is protected, the concept of discrimination also requires a definition of what discrimination *is*. Although computer scientists have advanced a variety of formal definitions, we rely on existing legal theories and, in particular, disparate impact doctrine, which scrutinizes disparities in selection rates that systematically disadvantage marginalized groups. A disparity in selection rates occurs when the rate of positive outcomes differs between groups—for example, when a model used for lending decisions approves a lower proportion of Black than white applicants . . . ." (emphasis in original) (footnote omitted)).

220. Kim, *Race-Aware Algorithms*, *supra* note 39, at 1557 ("Disparate impact theory is relevant to predictive algorithms because these tools may disproportionately screen out racial create a disparate impact, discrimination can exist when algorithms produce a disparate impact due to their outcome variables, even if their developers did not intend it. According to the statistical discrimination argument, mandating the application of fairness constraints is equivalent to prohibiting statistical discrimination through disparate impact doctrine. Because the choice of outcome variable is what generated the statistical bias, it is reasonable to require that statistical measures be taken to correct for such bias if the decision falls under disparate impact discrimination.

Disparate impact discrimination does not apply to all decisions, but avoiding a disparate impact matters for policy. To address proven biases such as those found in COMPAS, private providers of decision-assistance algorithms that are used by government agencies or the judiciary should be required to make reasonable efforts to apply fairness constraints in order to reduce such biases. These constraints could be applied by "permitting the use of protected traits (like race and sex) within the algorithm to determine what other traits will be used to predict the target variable (like recidivism)."[221]

Acknowledging that outcome variables are estimations for ideal decision criteria allows one to see that applying fairness constraints might improve the accuracy of the algorithm. Many are reluctant to entertain this approach because they believe that it is prohibited by antidiscrimination law in its prohibition of race-based classification.[222] However, as Deborah Hellman demonstrates, the use of protected traits in algorithms is likely legally permissible because "the use of race to determine what other factors to include within an algorithm" does not necessarily constitute disparate treatment.[223] Hellman's conclusion shows how the categories of disparate treatment and disparate impact are often seen as more rigid and obstacle-ridden than they actually are, and how they may serve instead as frameworks for improving algorithmic accuracy while lessening discrimination.[224] Moreover, because race *causes* the disparity,[225] even if the COMPAS model does not consider race, and even absent discriminatory intent, Black individuals do receive higher risk scores on the basis of their race—which means that they are treated differently.[226] Eliminating differential treatment should not be considered disparate treatment.

The gap between the relevant outcome and its measurement matters for determining who wins and who loses. False positives and false negatives are defined

---

minorities from employment opportunities, even if the employer did not intend to discriminate when adopting the tool."); *see also* Barocas & Selbst, *supra* note 42, at 701–12.

221. Hellman, *supra* note 182, at 818.

222. *See* Corbett-Davies et al., *Algorithmic Decision Making and the Cost of Fairness*, *supra* note 12, at 805 ("[E]xplicitly including race as an input feature raises legal and policy complications . . . .").

223. Hellman, *supra* note 182, at 848.

224. *Id.* at 848–49 ("This conclusion . . . matters conceptually because it demonstrates the way in which the categories of *disparate treatment* and *disparate impact* are less distinct and more porous than current legal doctrine acknowledges." (emphasis in original)).

225. *See supra* Section I.C.

226. Because the groups are treated in fact differently on the basis of race, it is just the intention requirement that is the obstacle for the racial disparity to constitute disparate treatment.

in terms of the predictor variable of arrests, and not in terms of the desired knowledge of recidivism. Therefore, displaying an increment of false positives as to rearrest data (an apparent loss in accuracy) *is not necessarily a bad thing*. An increment of false positives as to rearrest data would be a "cost to society" if arrest and crime tracked perfectly—but they do not. Because the socially valuable metric is the actual risk of committing a (violent) crime, an asymmetric false positive rate as to the rearrest outcome variable produced by a fairness adjustment, at least directionally, would correct for the racial bias in arrest data that the model amplifies.[227] This operation could, hypothetically, overcorrect or undercorrect, so it is helpful to have a sense of how biased arrest data might be with regards to actual criminality. If the application of a fairness constraint corrects by about the amount that one believes arrest data is racially biased, or if it undercorrects, then there is no social cost in terms of safety of applying a fairness constraint as it is usually stated.

Imagine one held the (mistaken) belief that COMPAS does not disadvantage a protected category by categorizing more Black individuals as high risk and, therefore, that training it with fair data would lead to lower predictive accuracy for the target.[228] In that case, whether one considers the COMPAS algorithm discriminatory would depend on whether there is a business justification, where the justification would be to avoid the cost of inaccuracy.

This justification would not extend to the private costs of improving the algorithm. In many circumstances, the law considers decisions as discriminatory even if a prediction is true because the disparate impact doctrine requires the decision-maker to adjust its expectations or provide alternatives.[229] Since algorithmic discrimination is often a disparate impact problem,[230] when disparate impact is applicable, one should expect the law to introduce some cost on decision-makers, like disparate impact antidiscrimination law does for other decisions.[231]

This leads to the normative point of how the costs of inaccuracy should be distributed. The costs of poor classification are currently borne by marginalized groups, who are disproportionately misevaluated as high-risk.[232] From a policy viewpoint, it would be an improvement to allocate those costs in a way that reduces the discrepancy between what is being measured and what ought to be estimated.

Shifting the cost of classification inaccuracy to model providers (or, in some circumstances, to decision-makers) through liability mechanisms would provide incentives to invest in improving model performance with regard to what ought to be estimated. Regulating the discriminatory outcomes of algorithms is thus likely to be welfare-increasing by reducing negative externalities to the populations being discriminated against—and, in the process, overcoming collective action problems that arise from large groups with few resources having to act against these decision systems. Moreover, like other aspects of antidiscrimination law, doing so would have desirable distributive purposes.[233] When there is a proven discriminatory outcome like there is in COMPAS, the provider of a decision-making algorithm should have a burden of proof on the adequacy of the outcome variable and data used.[234]

---

227. *See generally* Prince & Schwarcz, *supra* note 6.

228. *See supra* notes 142–144.

229. *See* Christine Jolls, Commentary, *Antidiscrimination and Accommodation*, 115 HARV. L. REV. 642, 697–99 (2001).

230. *See* Barocas & Selbst, *supra* note 42, at 701–12.

231. *See, e.g.*, Bradley v. Pizzaco of Neb., Inc., 939 F.2d 610 (8th Cir. 1991); Bradley v. Pizzaco of Neb., Inc., 7 F.3d 795 (8th Cir. 1993); Lanning v. Se. Pa. Transp. Auth., 181 F.3d 478 (3d Cir. 1999); *see also* Jolls, *supra* note 229, at 654–55; Wendy W. Williams, *Equality's Riddle: Pregnancy and the Equal Treatment/Special Treatment Debate*, 13 N.Y.U. REV. L. & SOC. CHANGE 325, 368 (1985).

232. *See supra* Sections II.B–II.C.

This argument extends to the very exercise of using AI to predict risk across crimes. The argument that COMPAS supporters often use, and that Northpointe itself has used, is the social utility of using data to predict future criminality.[235] If that social utility exists, and, historically, members of one race are arrested more for a specific type of crime, then the predictive operation is unhelpful because the prediction trained on historical data cannot capture changes in patterns of criminality. For some types of crimes, such as drug possession, the disparate rate of arrests is worse than for others.[236] Some categories of data, such as data on drug arrests, are so racially skewed that one should avoid using them to predict recidivism at all.

### *D. What Kind of Bias Is the COMPAS Bias?*

Our results relate to identifying the source of bias in algorithms like COMPAS.[237] In Table 6, as discussed, the first disparate impact ratio represents the bias in the dataset while the second ratio represents the bias in the AI output.[238] Because COMPAS uses a robust sampling of arrest data (where the sample resembles the population),[239] and because research on it provides no reason to suspect human biases

---

233. *See* John Gardner, *Discrimination as Injustice*, 16 OXFORD J. LEGAL STUD. 353, 355–56 (1996); John Gardner, *Liberals and Unlawful Discrimination*, 9 OXFORD J. LEGAL STUD. 1, 11 (1989).

234. *See* Black et al., *supra* note 14, at 90 ("[W]e argue that companies should have a duty to reasonably search for these alternatives rather than only being liable if they failed to adopt LDAs that they actually considered in the development process. This approach increases the likelihood of discovering LDAs because it creates incentives for developers, who are in the best position to discover them, to search for them in the first place." (footnote omitted)).

235. *See, e.g.*, TIM BRENNAN & ANGEL ILARRAZA, NORTHPOINTE, CONNECTING THE DOTS: SUPPORTING EVIDENCE-BASED SENTENCING DECISIONS WITH RISK-NEED-RESPONSIVITY PRINCIPLES 8 (2015), https://www.michbar.org/file/news/releases/archives17/Connecting-the-Dots-White-Paper.pdf [https://perma.cc/2UFX-9C9A] ("The implications of a risk score are clear. High-risk offenders – particularly violent and habitual offenders – should be given higher intensity treatment programming, more incarceration and supervision levels that are consistent with considerations of public safety and proportionality."); Adam Neufeld, *In Defense of Risk-Assessment Tools*, MARSHALL PROJECT (Oct. 22, 2017, 10:00 PM), https://www.themarshallproject.org/2017/10/22/in-defense-of-risk-assessment-tools [https://perma.cc/77BS-K3WN].

236. *See supra* notes 154–159.

237. *See supra* Section I.A.

238. *See supra* Section II.C.

239. Tim Brennan & William Dieterich, *Correctional Offender Management Profiles for Alternative Sanctions (COMPAS)*, *in* HANDBOOK OF RECIDIVISM RISK/NEEDS ASSESSMENT TOOLS (Jay P. Singh, Daryl G. Kroner, J. Stephen Wormith; Sarah L. Desmarais & Zachary encoded in the process (i.e., we have no reason to believe that it has a human problem in the coding or labeling stages), the source of bias in the COMPAS *dataset* is social bias. The COMPAS bias can be understood through a sociotechnical approach as a replication and amplification of a systemic bias within the criminal justice system. At its core, the COMPAS dataset bias reflects the historical, structural, and cultural inequalities embedded within these institutions. These biases can be traced back to factors like racial disparities in policing, socioeconomic inequalities, and biased sentencing practices. When adding AI, the bias can be perpetuated through feedback loops where past decisions inform future ones, creating a self-reinforcing cycle.

Treating the COMPAS bias as a sample bias might seem warranted at first sight because it stems from skewed data used to train the model. But it is the biased representative historical data, influenced by systemic inequalities, that leads to incorrect predictions and reinforces discriminatory outcomes. Treating it as a sample bias would focus solutions on addressing underlying data collection practices. However, our results show that doing so is insufficient: Better sampling would not fix the bias, and the model itself increases the bias in the dataset. Recognizing the COMPAS bias as a social bias shapes the way we think about reforming algorithms and mitigating algorithmic harm.

Identifying the COMPAS algorithm's bias as entirely a replication of social bias, however, misses the important fact that the algorithm makes that social bias worse. In the regression model we used, 79.42% of the output bias is driven by the COMPAS dataset and 20.58% by the model itself, [240] making it primarily, but not entirely, the third type of algorithmic bias in the categorization above.[241]

There are legal implications of mistakenly treating the COMPAS output bias as entirely social bias by ignoring the importance of measuring the right variable. Both sampling and social algorithmic biases can trigger disparate impact discrimination—although the first can be corrected by appropriate sampling, the second one cannot. Some believe that, if an AI model merely mirrors social biases, correcting it may lead to concerns of illegal affirmative action under current Supreme Court doctrine.[242] The issue being one where using AI *amplifies* a social bias rather than merely replicating it should eliminate affirmative action concerns that may make the fairness constraints illegal from a prohibited classification standpoint. First, because these fairness methods avoid classification.[243] Second, because the intervention, far from implementing affirmative action, corrects for an additional layer of inequality

---

Hamilton eds., 2017) ("For implementations in new agencies COMPAS typically uses a large norming (standardization) representative sample of the agencies' target population.").

240. *See supra* note 90; *infra* Appendix B, Table 6 (noting the 20.58% attributable to the model, where the remainder is 79.42%).

241. *See supra* notes 43–45.

242. *See* Jason R. Bent, *Is Algorithmic Affirmative Action Legal?*, 108 GEO. L.J. 803, 805 (2020) (considering, for example, whether weighting AI algorithms to account for bias in promotion decisions would violate affirmative action precedent); Daniel E. Ho & Alice Xiang, *Affirmative Algorithms: The Legal Grounds for Fairness as Awareness*, U. CHI. L. REV. ONLINE (Oct. 30, 2020), https://lawreviewblog.uchicago.edu/2020/10/30/aa-ho-xiang/ [https://perma.cc/6PGX-SPL6]. *See generally* Students for Fair Admissions, Inc. v. President & Fellows of Harvard Coll., 600 U.S. 181 (2023).

243. *See* Hellman, *supra* note 182.

that the system introduces: The counterfactual to using COMPAS (i.e., not using an AI system to predict risk) disadvantages minority groups less even under existing socioeconomic conditions.

Additionally, we find that racial bias is not the only bias present in COMPAS. Other variables, such as marital status and age, also causally influence the risk score.[244] These new findings carry policy implications on how AI should be used in the criminal justice system. First, these findings, at minimum, support frequent calls in favor of transparency for procedural purposes.[245] Second, if we continue to find new forms of discrimination in a dataset that has been analyzed extensively in the past, the criminal justice system should consider not making criminal justice decisions based on AI predictions of future criminality at all.[246]

From a sociotechnical perspective, the COMPAS bias is a symptom of a broader issue within the criminal justice system where certain groups are disproportionately targeted, arrested, and sentenced. Addressing the specific bias in COMPAS is desirable, but addressing its root causes requires systemic changes. Reform efforts would ideally include reevaluating policing methods and addressing economic disparities. A holistic approach that attempts to dismantle systemic biases once quantitative efforts like COMPAS provide evidence of them might help work toward a more equitable criminal justice system overall.

## CONCLUSION

This Article identifies the implications of using fair AI models to alleviate bias and discrimination in recidivism predictions. By empirically analyzing the COMPAS dataset using causal methods, it first confirms the existence of racial bias in both the training dataset and the AI prediction results. It then demonstrates that common AI models trained with a biased dataset *amplify* the bias—and they can do so by a meaningful margin. Then, it applies fair AI models to mitigate the racial bias and examines their relationship with model accuracy.

It is not necessarily the case that society pays a cost for fairness improvements in criminal justice decisions involving AI. When a fair AI model is used, false negatives (low-risk individuals mistakenly predicted as high risk) decrease and false positives (high-risk individuals mistakenly predicted as low risk) increase. This increment makes it seem like there is an accuracy loss, supporting the popular representation of a cost to pay for fairness. However, an increment in false positives is desirable because it corrects for the discrepancy between the outcome variable and the desired metric. There is a popular argument in policy, business, and academia that society "pays" the cost of AI fairness in criminal justice because future (racialized) criminals are released if decisions are made based on fairness-adjusted AI models' predictions. Our results reveal the flaw in such an argument.

---

244. *See infra* Appendix C.

245. *See, e.g.*, Danielle Keats Citron & Frank Pasquale, *The Scored Society: Due Process for Automated Predictions*, 89 WASH. L. REV. 1, 18–27 (2014).

246. *See, e.g.*, Ari Ezra Waldman, *Power, Process, and Automated Decision-Making*, 88 FORDHAM L. REV. 613, 617, 622–32 (2019); Alicia Solow-Niederman, YooJung Choi & Guy Van den Broeck, *The Institutional Life of Algorithmic Risk Assessment*, 34 BERKELEY TECH. L.J. 705, 710–18 (2019); Benjamin Eidelson, *Patterned Inequality, Compounding Injustice, and Algorithmic Prediction*, 1 AM. J.L. & EQUAL. 252, 252 (2021); Stevenson, *supra* note 26, at 330–31 ("If it is concerning that black defendants who do not recidivate are more likely to be labeled high risk than white defendants who do not recidivate (and there are plenty of reasons why this should be concerning!), then this calls into question the entire regime of using risk as a basis of restricting liberties, not simply actuarial risk assessment instruments.").

At a broader level, our finding that risk assessment algorithms like COMPAS—trained to maximize the accuracy of the prediction probability over a future arrest—magnify the racial biases in their datasets and have a causal relationship between race and higher risk scores contributes to two points. First, this finding illustrates that a reduction in outcome variable accuracy in COMPAS does not make a significant difference because the maximization of that outcome variable is itself racially biased.

Second, our results respond to a common question: Given that people are biased (in this case, producing biased arrests), why does it matter if AI systems are biased too? The answer is partly that, contrary to human decision-makers, AI algorithms operate at scale so, while human bias is in itself concerning, when integrated into algorithms, it becomes automated. But, perhaps more importantly, we show that AI processes can magnify, not just encode and replicate, racial bias. The choice of outcome variables in AI matters because these variables can amplify social biases in the database. This mechanism can increase discrimination and injustice. More broadly, besides providing further basis for adjusting recidivism prediction models with fairness constraints, our results can be used to support calls for ceasing the use of such models altogether—i.e., doing away with automated risk assessment in criminal justice.

## APPENDIX A: DATASET

The dataset contains the characteristics of each individual and the risk score (decile score) that COMPAS produces. The total number of observations is 60,789. The primary dependent variable of interest in this study is a binary variable (Race_Black), which indicates whether an individual is Black. The independent variable of interest is $DecileScore$, which is the output of the COMPAS system for each individual. The score ranges from 1 to 10, where 1 corresponds to a prediction that the individual has the lowest risk to reoffend while 10 corresponds to a prediction that the individual has the highest risk to reoffend. For classification tasks, we transform this variable into a binary variable that takes the value 1 for observations with a decile score higher than or equal to 5 and 0 otherwise. A summary statistics table of the COMPAS dataset is provided in Table 2.

**Table 2:** Summary Statistics of Variables in the COMPAS Dataset

| Variables | Mean | Min. | Max. | Std. dev. |
| --- | --- | --- | --- | --- |
| Race_Black | 0.4450 | 0 | 1 | 0.4970 |
| DecileScore | 3.5751 | 1 | 10 | 2.6159 |
| Agency_Text_Broward County | 0.002023 | 0 | 1 | 0.04493 |
| Agency_Text_DRRD | 0.009277 | 0 | 1 | 0.09587 |
| Agency_Text_PRETRIAL | 0.6756 | 0 | 1 | 0.4681 |
| Agency_Text_Probation | 0.3131 | 0 | 1 | 0.4637 |
| Sex_Code | 0.7809 | 0 | 1 | 0.4136 |
| DateOfBirth_yr (19xx) | 79.3246 | 18 | 98 | 12.1710 |
| ScaleSet_ID | 0.9639 | 0 | 1 | 0.1866 |
| AssessmentType_Code | 0.9227 | 0 | 1 | 0.2671 |
| Language | 0.9959 | 0 | 1 | 0.06425 |
| LegalStatus_Conditional Release | 0.006875 | 0 | 1 | 0.08263 |
| LegalStatus_Deferred Sentencing | 0.000197 | 0 | 1 | 0.01405 |
| LegalStatus_Other | 0.07153 | 0 | 1 | 0.083 |
| LegalStatus_Parole Violator | 0.000296 | 0 | 1 | 0.01720 |
| LegalStatus_Post Sentence | 0.3013 | 0 | 1 | 0.4588 |
| LegalStatus_Pretrial | 0.6177 | 0 | 1 | 0.4859 |
| LegalStatus_Probation Violator | 0.002072 | 0 | 1 | 0.04548 |
| CustodyStatus_Jail Inmate | 0.4953 | 0 | 1 | 0.5000 |
| CustodyStatus_Parole | 0.000345 | 0 | 1 | 0.01858 |
| CustodyStatus_Pretrial Defendant | 0.1257 | 0 | 1 | 0.3315 |
| CustodyStatus_Prison Inmate | 0.000247 | 0 | 1 | 0.01571 |
| CustodyStatus_Probation | 0.3773 | 0 | 1 | 0.4847 |
| CustodyStatus_Residential Program | 0.001036 | 0 | 1 | 0.03217 |
| MaritalStatus_Divorced | 0.06377 | 0 | 1 | 0.2443 |
| MaritalStatus_Married | 0.1343 | 0 | 1 | 0.3409 |
| MaritalStatus_Separated | 0.02964 | 0 | 1 | 0.1696 |
| MaritalStatus_Significant Other | 0.02069 | 0 | 1 | 0.1424 |
| MaritalStatus_Single | 0.7417 | 0 | 1 | 0.4377 |
| MaritalStatus_Unknown | 0.003503 | 0 | 1 | 0.05909 |
| MaritalStatus_Widowed | 0.006415 | 0 | 1 | 0.07984 |
| Screening_Date_yr (19xx) | 13.4669 | 4 | 14 | 0.5850 |
| RecSupervisionLevel | 1.6304 | 1 | 4 | 0.9444 |
| DisplayText_Risk of Failure to Appear | 0.3336 | 0 | 1 | 0.4715 |
| DisplayText_Risk of Recidivism | 0.3330 | 0 | 1 | 0.4713 |
| DisplayText_Risk of Violence | 0.3334 | 0 | 1 | 0.4714 |

## APPENDIX B: RACIAL BIAS RESULTS

We mitigate the discovered bias using the GridSearch reduction approach, which is available as part of a popular open-source library package containing techniques to detect and mitigate bias in machine learning models.[247] That package, called Fairlearn, allows developers to assess AI systems' fairness to mitigate observed unfairness issues and contains mitigation algorithms as well as different metrics for model assessment.[248] The idea behind using the Fairlearn package is to examine the accuracy of two groups (e.g., white individuals and Black individuals) to see whether an algorithm tends to predict outcomes for one group better than for the other.

FACT can use multiple methods to assess similarities between the groups, and the similarities can be based on more than one variable. In our empirical analyses, we use three matching methods to assess the similarities between Black individuals and other individuals: Nearest Neighbor (NNM), Nearest Neighbor with Propensity Caliper (NNMPC), and Mahalanobis Metric Matching with Propensity Caliper (MMMPC). These techniques are available as part of publicly available packages.[249]

**Table 3:** Evidence of Racial Bias with Causal Inference (Black Individuals)

| Methods | Estimate ($\gamma$) | Std. Error | t-value | p-value |
|---|---|---|---|---|
| FACE | 0.1901 | 0.01483 | 12.8184 | < 0.001 |
| FACT-NNM | 0.1590 | 0.01632 | 9.7405 | < 0.001 |
| FACT-NNMPC | 0.1242 | 0.01748 | 7.1079 | < 0.001 |
| FACT-MMMPC | 0.1755 | 0.01637 | 10.7222 | < 0.001 |
| Causal Forest | 0.1242 | 0.01631 | NA | NA |

**Table 3b:** Evidence of Racial Bias with Causal Inference (Hispanic Individuals)

| Methods | Estimate |
|---|---|
| FACE | 0.0931 |
| FACT-NNM | 0.1980 |
| FACT-NNMPC | 0.0819 |
| FACT-MMMPC | 0.0974 |
| Causal Forest | 0.0528 |

Table 3b finds similar results for Hispanic individuals as for Black individuals (with a lower p-value).

---

247. BIRD ET AL., *supra* note 106, at 5–6. Note that it works throughout the AI application life cycle.

248. *Fairness in Machine Learning*, FAIRLEARN, https://fairlearn.org/main/user_guide/fairness_in_machine_learning.html#fairness-assessment-and-unfairness-mitigation [https://perma.cc/3EKZ-7KSW].

249. *See* BIRD ET AL., *supra* note 106; Bellamy et al., *supra* note 10.

**Table 4:** Causal Forest Goodness of Fit (Black Individuals)

| Methods | Estimate | Std. | t-value | p-value |
|---|---|---|---|---|
| Mean Forest Prediction | 0.9826 | 0.1294 | 7.5928 | < 0.001 |
| Differential Forest | 0.5955 | 0.06814 | 8.7396 | < 0.001 |

Note from the results that the estimation from causal forest does not directly include statistical properties. To evaluate the statistical significance of this estimation, we use the test_calibration() function in the causal forest package,[250] which outputs mean.forest.prediction and differential.forest.prediction. Both values, which represent the goodness of fit of forests developed by causal forest,[251] are reported in Table 3. The coefficient of 1 for the mean forest prediction indicates that the mean forest prediction is correct. Meanwhile, the coefficient of 1 for differential forest prediction indicates that the forests have captured heterogeneity in the underlying signal. We find the mean forest prediction to be close to 1, and it is statistically significant ($p < 0.001$). Meanwhile, the differential forest prediction is 0.5955, and it is statistically significant ($p < 0.001$). As a result, the coefficient that causal forest produces is likely correct, and the forests reasonably capture the heterogeneity in the underlying signal.

**Table 4b:** Causal Forest Goodness of Fit (Hispanic Individuals)

| Methods | Estimate ($\gamma$ ) |
|---|---|
| Mean Forest Prediction | 0.9443 |
| Differential Forest Prediction | 0. 8965 |

Table 4b also finds similar results for Hispanic individuals as for Black individuals.

**Table 5:** Evidence of Bias with Discrepancy in Prediction Accuracy (Black Individuals)

| Category | Model Accuracy |
|---|---|
| Black Individuals | 0.8132 |
| Others | 0.8958 |
| Overall Model Accuracy | 0.8592 |

We test the statistical significance of this discrepancy using the permutation test[252]: We perform a two-sided permutation test and set the number of iterations to

250. Tibshirani et al., *supra* note 69.
251. *Evaluating a Causal Forest Fit*, GRF LABS, https://grf-labs.github.io/grf/articles/diagnostics.html [https://perma.cc/LW9N-99GQ].
252. PHILLIP GOOD, PERMUTATION TESTS: A PRACTICAL GUIDE TO RESAMPLING METHODS FOR TESTING HYPOTHESES (2013).

10,000 so that we obtain the p-value with four decimal places. We find that the discrepancy under the logistics regression model has a p-value < 0.001.

**Table 5b:** Evidence of Bias with Discrepancy in Prediction Accuracy (Hispanic Individuals)

| Category | Model Accuracy |
|---|---|
| Hispanic | 0.8260 |
| Others | 0.8800 |
| Overall Model Accuracy | 0.8485 |

Table 5b also finds similar results for Hispanic individuals as for Black individuals.

**Table 6:** Bias from AI Model Versus Bias in Training Dataset (Black Individuals)

| Category | Disparate Impact Ratio Accuracy |
|---|---|
| Original Dataset | 0.4821 |
| AI Predictions | 0.3755 |

**Table 6b:** Bias from AI Model Versus Bias in Training Dataset (Hispanic Individuals)

| Category | Disparate Impact Ratio Accuracy |
|---|---|
| Original Dataset | 0.5912 |
| AI Predictions | 0.5159 |

Table 6b finds similar results for Hispanic individuals as for Black individuals (with a lower p-value).

**Table 7:** Accuracy Difference (Grid Search with DP) (Black Individuals)

| Groups | Model 0 | Model 29 | Percentage Change |
|---|---|---|---|
| Black | 0.7601 | 0.5250 | -30.93% |
| Others | 0.8186 | 0.7817 | -4.51% |

**Table 7b:** Accuracy Difference (Grid Search with DP) (Hispanic Individuals)

| Groups | Model 0 | Model 29 | Percentage Change |
|---|---|---|---|
| Hispanic | 0.7742 | 0.6032 | -22.08% |
| Others | 0.8109 | 0.7777 | -4.09% |

Table 7b finds similar results for Hispanic individuals as for Black individuals (with a lower p-value).

**Table 8:** Confusion Matrix (Gridsearch with DP) (Black Individuals)

| Counts | Model 0 (Black) | Model 29 (Black) | Model 0 (Others) | Model 29 (Others) |
|---|---|---|---|---|
| True Positive | 296 | 357 | 417 | 443 |
| True Negative | 155 | 12 | 55 | 8 |
| False Positive | 107 | 250 | 124 | 171 |
| False Negative | 63 | 2 | 27 | 1 |

**Table 8b:** Confusion Matrix (Gridsearch with DP) (Hispanic Individuals)

| Counts | Model 0 (Hispanic) | Model 29 (Hispanic) | Model 0 (Others) | Model 29 (Others) |
|---|---|---|---|---|
| True Positive | 376 | 436 | 337 | 356 |
| True Negative | 162 | 37 | 48 | 8 |
| False Positive | 126 | 251 | 105 | 145 |
| False Negative | 67 | 7 | 23 | 4 |

Table 8b finds similar results for Hispanic individuals as for Black individuals (with a lower p-value).

**Table 9:** Accuracy Difference (Grid Search with EO) (Black Individuals)

| Groups | Model 0 | Model 8 | Percentage Change |
|---|---|---|---|
| Black | 0.7375 | 0.6924 | -6.11% |
| Others | 0.8042 | 0.8170 | 1.5968% |

**Table 9b:** Accuracy Difference (Grid Search with EO) (Hispanic Individuals)

| Groups | Model 0 | Model 8 | Percentage Change |
|---|---|---|---|
| Hispanic | 0.7400 | 0.7277 | -1.66% |
| Others | 0.8031 | 0.8206 | 2.179% |

Table 9b finds similar results for Hispanic individuals as for Black individuals (with a lower p-value).

**Table 10:** Confusion Matrix (Gridsearch with EO) (Black Individuals)

| Counts | Model 0 | Model 8 | Model 0 | Model 8 |
|---|---|---|---|---|
| True Positive | 263 | 332 | 399 | 432 |
| True Negative | 200 | 129 | 90 | 51 |
| False Positive | 62 | 133 | 89 | 128 |
| False Negative | 96 | 27 | 45 | 12 |

**Table 10b:** Confusion Matrix (Gridsearch with EO) (Hispanic Individuals)

| Counts | Model 0 | Model 8 | Model 0 | Model 8 |
|---|---|---|---|---|
| True Positive | 334 | 405 | 315 | 345 |
| True Negative | 221 | 149 | 82 | 51 |
| False Positive | 67 | 139 | 71 | 102 |
| False Negative | 109 | 38 | 45 | 15 |

Table 10b finds similar results for Hispanic individuals as for Black individuals (with a lower p-value).

## APPENDIX C: BIAS FOR AGE AND MARITAL STATUS

We perform further analysis on COMPAS to determine whether there are other biases of concern besides the well-known racial bias. We use the methodologies described above. We find two biases: bias on marital status (more false positives for single individuals) and on age (more false positives for older individuals).

The second variable we analyze is the marital status of each individual in the dataset. This variable is commonly found to induce bias in employment.[253] Here, we ask whether being married affects the risk score in the prediction of recidivism risk. Results from Table 11 demonstrate a bias on marital status. Married individuals are given 30.96% to 32.58% lower risk scores than unmarried individuals *who are otherwise similar*.

**Table 11:** Evidence of Bias with Causal Inferences (Marital Status)

| Methods | Estimate ($\gamma$) | Std. Error | t-value | p-value |
|---|---|---|---|---|
| FACE | -0.3101 | 0.02410 | -12.8690 | < 0.001 |
| FACT-NNM | -0.3243 | 0.02742 | -11.8273 | < 0.001 |
| FACT-NNMPC | -0.3214 | 0.03218 | -9.9896 | < 0.001 |
| FACT-MMMPC | -0.3258 | 0.02782 | -11.7110 | < 0.001 |
| Causal Forest | -0.3096 | 0.02879 | NA | NA |
| Mean Forest Prediction | 1.03843 | 0.1236 | 8.4025 | < 0.001 |
| Differential Forest Prediction | 0.8283 | 0.09136 | 9.06599 | < 0.001 |

Lastly, we analyze the potential bias induced by age. Age discrimination is one of the most common forms of discrimination in the United States.[254] Here, we

253. *See, e.g.*, Alexander H. Jordan & Emily M. Zitek, *Marital Status Bias in Perceptions of Employees*, 34 BASIC & APPLIED SOC. PSYCH. 474 (2012); Joel T. Nadler & Katie M. Kufahl, *Marital Status, Gender, and Sexual Orientation: Implications for Employment Hiring Decisions*, 1 PSYCH. SEXUAL ORIENTATION & GENDER DIVERSITY 270 (2014).

254. *See, e.g.*, David Neumark, *Age Discrimination Legislation in the United States*, 21 CONTEMP. ECON. POL'Y 297 (2003).

examine whether age causally impacts the risk score in the prediction of recidivism risk. Note that because age is a nonbinary variable, we cannot apply FACE and FACT for the analysis. Therefore, only results from causal forest are reported in Table 12. Causal forest estimates that younger people receive an approximately 0.93% increment in the risk score when other characteristics are equivalent. Results from two additional correlational methods, Mean Forest Prediction and Differential Forest Prediction, strongly support the validity of this finding.

**Table 12:** Evidence of Bias with Causal Inferences (Age)

| Methods | Estimate ($\gamma$) | Std. Error | t-value | p-value |
|---|---|---|---|---|
| Causal Forest | 0.009306 | 0.0008087 | NA | NA |
| Mean Forest Prediction | 1.06001 | 0.3684 | 2.8776 | 0.002 |
| Differential Forest Prediction | 0.9879 | 0.01308 | 75.5345 | < 0.001 |

Why does this matter? These biases are common in other areas, such as employment.[255] They are consistent with biases found in judges.[256] Both have problematic disparate impact implications.

COMPAS's bias comes primarily from bias in rearrest data, but using AI systems such as COMPAS not only perpetuates but also grows such bias. This second finding is significant, considering that it uncovers further biases in one of the systems that has been explored in the literature more extensively. It indicates that there could be more problematic biases, which will also vary across systems being developed.[257]

This second finding confirms arguments in legal scholarship that while racial bias is the most problematic issue in the use of automatic risk assessment in criminal justice, there is a larger problem with the very method that filters into other protected categories in ways that are difficult to predict and detect.

---

255. Pauline T. Kim & Sharion Scott, *Discrimination in Online Employment Recruiting*, 63 ST. LOUIS U. L.J. 93 (2018); Kim, *Data-Driven Discrimination at Work*, *supra* note 6; Nadler & Kufahl, *supra* note 253; Jordan & Zitek, *supra* note 253; Neumark, *supra* note 254.

256. Megan T. Stevenson & Jennifer L. Doleac, *Algorithmic Risk Assessment in the Hands of Humans*, *in* IZA DISCUSSION PAPER SERIES NO. 12853, 1 (2019).

257. *See, e.g.*, Malathi A. & Baboo, *supra* note 2.